\documentclass[12pt]{iopart}
\usepackage{iopams}

\expandafter\let\csname equation*\endcsname\relax

\expandafter\let\csname endequation*\endcsname\relax

\usepackage{amsmath}
\usepackage{amsfonts}
\usepackage{amssymb}
\usepackage{graphicx}
\usepackage{epsf}
\usepackage{color}
\usepackage{wasysym}
\usepackage{color}
\usepackage{bbm}

\begin{document}

\title{Coherent back and forward scattering peaks in the quantum kicked rotor}

\author{G.~Lemari\'e $^1$, C.~A.~M\"uller $^{1,2}$, D.~Gu\'ery-Odelin $^3$, and C.~Miniatura $^{4,5,6,7,8}$}
\address{$^1$ Laboratoire de Physique Th\'eorique, IRSAMC, Universit\'e de Toulouse, CNRS, UPS, France\\
$^2$ Fachbereich Physik, Universit\"at Konstanz, 78457 Konstanz, Germany\\
$^3$ Laboratoire Collisions, Agr\'egats, R\'eactivit\'e, IRSAMC, Universit\'e de Toulouse, CNRS, UPS, France\\
$^4$ MajuLab, CNRS-UNS-NUS-NTU International Joint Research Unit, UMI 
3654, Singapore\\
$^5$ Centre for Quantum Technologies, National University of Singapore, 
3 Science Drive 2, Singapore 117543, Singapore\\
$^6$ Department of Physics, National University of Singapore, 2 Science 
Drive 3, Singapore 117542, Singapore\\
$^7$ School of Physical and Mathematical Sciences, Nanyang Technological University, Singapore 637371, Singapore\\
$^8$ Universit\'{e} C\^ote d'Azur, CNRS, INLN; 1361 route des Lucioles, 06560 Valbonne, France}

\date{\today}

\begin{abstract}
We propose and analyze an experimental scheme using the quantum kicked rotor to observe the newly-predicted coherent forward scattering peak together with its long-known twin brother, the coherent backscattering peak. Contrary to coherent backscattering, which arises already under weak-localization conditions, coherent forward scattering is 
only triggered by Anderson or strong localization. 
So far, coherent forward scattering has not been observed in conservative systems with elastic scattering by spatial disorder. We propose to turn to the quantum kicked rotor, which has a long and succesful history as an accurate experimental platform to observe dynamical localization, i.e., Anderson localization in momentum space. 
We analyze the coherent forward scattering effect for the quantum kicked rotor by extensive numerical simulations, both in the orthogonal and unitary class of disordered quantum systems, and show that an experimental realization involving phase-space rotation techniques is within reach of state-of-the-art cold-atom experiments.
\end{abstract}

\maketitle

\section{Introduction}

As is now well known, interference is a crucial ingredient to fully understand transport properties of waves propagating in media with quenched spatial disorder. Phase coherence is responsible for non-trivial weak localization corrections to the large-scale diffusion predicted by Drude-Boltzmann theory \cite{chandrabook1960, mesobook2007}. Ultimately, interference can even suppress transport entirely, a phenomenon known as strong (or Anderson) localization \cite{anderson58, kinnon1993}. Localization is the rule for large enough 1D and 2D systems, but a genuine, disorder-driven metal-insulator transition, the Anderson transition (AT), occurs in 3D \cite{gang4, evers2008}. Over the past fifty years, numerous experiments have revealed the delicate intricacies of Anderson localization 
\cite{chabanov2000, schwartz2007, billy2008, roati2008, hu2008, kondov2011, jendrzejewski2012loca3D, sperling2016}.

In the context of cold atoms, localization experiments so far have monitored the real-space expansion of a wave packet \cite{modugno2010}. The atomic cloud first spreads diffusively, then slows down and finally comes to a halt (vanishing diffusion). The wings of the stationary density profiles then show an exponential decay as a function of position. However the critical properties the 3D AT remain difficult to analyze by only using these real-space observables \cite{MuellerShapiroComment2014,delande2014mobility,semeghini2015measurement}. Indeed, the main challenge is to circumvent the large disorder-induced energy broadening of the initial state \cite{trappe2015semiclassical,prat2016semiclassical} and to select a sufficiently narrow energy window to avoid blurring the energy dependence of the diffusion constant and of the localization length near the critical point \cite{MuellerDelandeShapiro2016}. Though energy-filtering methods have been proposed for cold atoms \cite{dao2007}, their successful experimental implementation is still lacking. 

Recent theoretical works have proposed instead to study the localization dynamics in \textit{reciprocal} space, \textit{i.e.} to start from a narrow initial wave packet centered at non-zero momentum ${\bf k}_0$ (ideally a plane wave) and to monitor the time evolution of the disorder-averaged momentum distribution. It has been shown that interference effects give rise to non-trivial structures in momentum space: two narrow peaks emerge on top of an otherwise isotropic (diffusive) background \cite{cherroret2012coherent, karpiuk2012coherent, lee2014dynamics, ghosh2014cfs2D, ghosh2015cbs3d, ghosh2016cfs3d}. On a time scale set by the elastic scattering time, one observes the rapid emergence of a 
coherent peak centered at $-{\bf k}_0$ \cite{cherroret2012coherent}. This peak is associated with the paradigmatic coherent backscattering (CBS) effect \cite{wiersma95} and has been recently observed with cold atoms \cite{jendrzejewski2012coherent, labeyrie2012enhanced}. 
At longer times, theoretical arguments predict the emergence of a second peak centered at ${\bf k}_0$, the coherent forward scattering (CFS) peak, on a time scale given by the Heisenberg time associated to a localization volume \cite{karpiuk2012coherent, lee2014dynamics, ghosh2014cfs2D, ghosh2016cfs3d}. 
For time reversal symmetric systems, the CBS/CFS peaks settle to a twin structure in the localized regime, and both can be used to extract the critical properties of the 3D AT \cite{ghosh2015cbs3d, ghosh2016cfs3d} as exemplified by \cite{cobus2016} for CBS. It is worth noticing that, contrary to the CBS peak that disappears when time reversal symmetry is broken, the CFS peak is robust and exists in other symmetry classes \cite{micklitz2014strong}. The CFS effect is thus a genuine marker of Anderson localization in the bulk. Despite extensive and accurate numerical studies, an experimental observation of the CFS peak is still lacking. Here again, the implementation of reliable energy-filtering methods remains an experimental bottleneck. 

With this article, we suggest to circumvent this problem by considering another system, the quantum kicked rotor (QKR). This system has been extensively studied in the framework of quantum chaos \cite{haake2010}. From an experimental point of view, it has allowed the study of interference effects mimicking those observed with spatial disorder in a very controlled way \cite{garreau2016quantum}. For example, dynamical localization, the equivalent of 1D Anderson localization in momentum space, has been observed with a simple magneto-optical trap already back in the nineties \cite{moore94, moore1995atom}. More recently, the 3D AT has been fully addressed, including a measurement of the critical exponents \cite{chabe2008experimental, lemarie2009, lemarie2010, lopez2012experimental, lopez2013phase}.  In fact, theoretical and experimental studies are arguably more convenient with the QKR than with other systems (see e.g. \cite{dahlhaus2011quantum, edge2012metallic, manai2015experimental, hainaut2016return}). One of the many reasons is that characteristic parameters, like the mean free path, or the localization properties \textit{do not depend} on (quasi-)energy. We thus propose to use the QKR for an experimental observation of the CBS/CFS peaks.

In the following, we briefly present our theoretical model, introduce the CBS/CFS peaks for the QKR, analyze their properties, and propose a realistic experimental route to their observation, all of this both in the orthogonal and unitary universality class.  We then briefly conclude with some perspectives and future work.

\section{Kicked Rotor Model in a nutshell}

The kicked rotor (KR) is a paradigmatic model system for classical and quantum chaos \cite{haake2010}. The corresponding Hamiltonian serving our purposes reads
\begin{equation}\label{H1}
 \mathcal{H}= \frac{P^2}{2m} - V_0 \cos (kX) \sum_n \delta(\mathcal T/T_0-n), 
\end{equation}
and has been realized experimentally using cold atoms exposed to light pulses \cite{moore1995atom}. It describes a particle with mass $m$ on a line 
that is periodically kicked with period $T_0$ by a sinusoidal 
potential of strength $V_0$ and spatial period $2\pi/k$. For mathematical convenience, the periodic train of kicks is idealized in Eq.~\eqref{H1} as a series of delta-functions (Dirac comb). 

\subsection{Classical dynamics and chaos}

Using $k^{-1}$, $T_0$, $p_0=m/(kT_0)$, $E_0=p_0^2/m$ as space, time, momentum and energy units, the classical dynamics is conveniently described by the dimensionless Hamiltonian $H= \mathcal{H}/E_0$, 
\begin{equation}\label{eq:HKR}
H = \frac{p^2}{2} - K \cos x  \sum_n \delta(t-n), 
\end{equation}
featuring the dimensionless variables $x=kX$, $t=\mathcal T/T_0$, $p=P/p_0$ and the stochasticity parameter $K=V_0/E_0$. This system can be reduced to Chirikov's standard map \cite{chirikov1979} and exhibits a transition to chaos above $K_c \approx 0.97$ \cite{greene1979}. Classical transport
is then described by a pseudo-random walk leading to an unbounded Brownian motion in momentum space. A small ``drop'' of initial conditions peaked around $p=0$ spreads diffusively as $\langle p^2(t)\rangle = 2D_{cl}t$ with a classical diffusion constant $D_{cl} \approx K^2/4$ for $K> 4$ \cite{mackay1984}. 

\subsection{Dynamical localization}

The quantum Hamiltonian is obtained by the usual canonical procedure $P \to -i\hbar \partial_X$, which translates into $p \to -i\hbar_e \partial_x$. It introduces an additional dimensionless constant, the effective Planck constant $\hbar_e= \hbar/S_{cl}$ where $S_{cl}= p_0/k = E_0T_0$ is the classical action associated with the system. Whereas the classical phase space dynamics is only governed by $K$, the quantum dynamics depends on both $K$ and $\hbar_e$ (the semi-classical regime is defined by $\hbar_e \ll 1$). In the cold atom community, an important energy scale is the recoil energy $E_R= \hbar^2k^2/(2m)$. It is easy to see that $K= s \hbar_e^2/4$ where $s= 2V_0/E_R$ is the dimensionless lattice depth of the kick potential. It is interesting to note that one can control independently the classical parameter $K$ and the quantum parameter $\hbar_e$, for instance by varying $T_0$ while maintaining $V_0T_0^2$ constant.

The quantum behavior of the KR stands in marked contrast with the classical behavior: the classical diffusive transport freezes after a characteristic time $\tau$ known as the Heisenberg time (or break time). This is the hallmark of the dynamical localization phenomenon, i.e. Anderson localization in momentum space. Then, the expansion of a wave packet, initially peaked around $ p = 0$, saturates to a stationary exponential distribution, as observed experimentally \cite{moore1995atom}.

\subsection{Quantum dynamics {\it via} stroboscopic quantum maps}
\label{quantumstroboscopics.sec}

Like its classical counterpart, the quantum dynamics is best captured by using successive snapshots of the system right after each kick. This stroboscopic movie is generated by iterating the evolution operator $U$ over one period, 
\begin{equation}
 U = U_x U_p = e^{i K \cos x/\hbar_e} e^{-i p^2/ 2 \hbar_e} \; ,
\end{equation}
which is itself the product of the free evolution operator $U_p$ and the kick operator $U_x$. Because the potential is space-periodic with period $1$, one can expand $U_x= \sum_n f_n(K/\hbar_e) \, e^{inx}$ in Fourier components. Using the generating function of Bessel functions, one finds $f_n(K/\hbar_e)  = i^{n} \, J_{n}(K/\hbar_e)$ where the $J_n$ are Bessel functions of the first kind \cite{abrastegun1972}. According to Bloch's theorem, any momentum state can be expressed as $|p\rangle = |(l+\beta)\hbar_e\rangle \equiv | l, \beta\rangle$ where $l$ is an integer and $ \beta \in [0,1) $ is the quasi-momentum. The resolution of identity then reads
\begin{equation}
\sum_{l\in\mathbbm{Z}} \int_0^1 d\beta \, |l,\beta\rangle\langle l, \beta| = \mathbbm{1}\; ,
\end{equation}
and the matrix elements of $U$ in momentum space are 
\begin{equation}
\langle l' \beta' |U| l, \beta\rangle = e^{-i\alpha_l} f_{l'-l}(K/\hbar_e) \, \delta(\beta'-\beta),
\end{equation}
with $\alpha_l = \hbar_e(l+\beta)^2/2$. We find that the stroboscopic quantum map $U$ {\it conserves} the quasi-momentum $\beta$. As one can see, the kick operator $U_x$ plays
the role of a hopping amplitude coupling the momentum states $| l, \beta\rangle$ to their ``lattice neighbors" $| l', \beta\rangle$, the hopping range being essentially restricted by the exponential decrease of the Bessel function to $ \vert l' - l \vert \lesssim K / \hbar_e $ \cite{abrastegun1972}. We see that $ K / \hbar_e \equiv \ell_s$ plays the role of an effective scattering mean free path.
The phase factors $ e^{-i \alpha_l} $ associated with the free evolution operator play the role of the random on-site energies of the Anderson model, the average over disorder realizations being replaced here by an average over initial quasi-momenta $ \beta $. They have indeed a pseudo-random character: when $ \hbar_e $ is incommensurate with $ 2 \pi $, the free evolution phases $\alpha_l$ are uniformly distributed on the circle \cite{birkhoff1931}. This means that, in contrast to other systems where the amount of disorder can be varied, the QKR system always shows ``maximal disorder", even if, at the same time, the scattering mean free path can be changed at will. As an approximation (random QKR model), one can thus simply forget about the exact expression of the $\alpha_l$, consider them as true random numbers uniformly distributed over $ [0, 2 \pi]$ and average over them. 

\subsection{Localization length and Heisenberg time}
From a mathematical point of view, statements about the localization properties of the Floquet eigenstates $|\phi_i\rangle$ of the evolution operator over one period 
\begin{equation}
U|\phi_i\rangle = e^{i \omega_i} |\phi_i\rangle
\end{equation}
with (real) quasi-energies $\omega_i \in [0,2\pi)$, can be related to theorems about products of unimodular random matrices and their Lyapunov exponents \cite{furstenberg1960, oseledec1968, haake2010}. From a physical point of view, Fishman {\it et al.} found an important connection between the QKR and a 1D Anderson model with quasi-random onsite disorder \cite{fishman1982}. Using field-theoretic methods, it was later shown that the correspondence with quasi-1D wires is exact \cite{altland1996}. In fact, simple and appealing theoretical arguments give the expected expression of the localization length \cite{shepe1987}. In the direct lattice space labelled by the momentum integers $n$, the diffusion constant is $D_{cl}/\hbar^2_e$. Since the diffusive growth is stopped after the break time $\tau$, it means that the kicks can only excite a finite number $\xi$ of lattice states, which identifies with the localization length in the direct space. Since disorder is maximal, the associated quasi-energies $\omega_i$ are uniformly distributed over $2 \pi$ and their mean level spacing within the localization volume $\xi$ is thus  $\delta E   \sim
2\pi/\xi $. Diffusion will continue until the discreteness of the spectrum is resolved around the time $\tau \sim 2\pi/\delta E \sim \xi$. The number of levels effectively involved is then $\sqrt{2D_{cl}\tau}/\hbar_e \sim \xi$, leading to the scaling relations \cite{shepe1987, fishman1989}
\begin{equation}
\label{eq:locaQKR}
\xi  \sim \tau \sim D_{cl}/\hbar^2_e \sim \frac{K^2}{4\hbar^2_e} \sim \ell_s^2/4
\end{equation}
The exact numerical prefactors cannot be determined from such heuristic arguments but these results show that the QKR is similar to a quasi-1D disordered system with $N_\perp \sim  \ell_s$ transverse channels. It is noteworthy that $\xi$, $\tau$ and $\ell_s$ {\it do not depend} on energy. This salient feature is a key advantage of the QKR over other disordered systems.

\subsection{Orthogonal and unitary class} 

As is well known, the localization properties of a system depend on its symmetry properties \cite{evers2008}. An important symmetry is time reversal (TR) with the twist that, compared to usual disordered systems, space and momentum exchange their role for the QKR. This implies that the TR symmetry is in fact here given by $t\to -t$, $x \to -x$ and $p\to p$, that is by the usual time reversal symmetry followed by space inversion (see \cite{hainaut2016return}). As one can readily see, the QKR Hamiltonian \eqref{eq:HKR} is TR symmetric in this sense and is said to belong to the orthogonal class \cite{mehta2004}. This also means that TR for the QKR can be conveniently broken by breaking space inversion, in which case the system falls into the unitary class. Following \cite{thaha1993symmetry}, this can be done by replacing the kick potential $V_{\text{kick}}(x)=\cos x $ in \eqref{eq:HKR} by a bichromatic superlattice \cite{weitz06,bloch08}
\begin{equation}
\label{eq:kickUC}
 \tilde{V}_{\text{kick}}(x)=\left[\cos\left(\frac{\pi q}{2}\right) \cos x+\frac{1}{2} \sin\left(\frac{\pi q}{2}\right) \sin(2x)\right],
\end{equation}
with a parameter taken to be $q=0.5$ in our numerical simulations. In the following we theoretically investigate the CFS and CBS properties of the QKR both for the orthogonal and unitary classes, leaving other possible symmetry classes like the symplectic one for future studies. 

The localization length in the unitary class is known to be twice the localization length in the orthogonal class, at fixed diffusion constant ({\it i.e}. at  fixed $K$ and $\hbar_e$ here) \cite{efetov1983,pichard1990,thaha1993symmetry}:

\begin{equation}
\label{eq:locaQKR-U}
\xi_u = 2\xi_o.
\end{equation}
Since we will study below CFS and CBS in both the unitary and orthogonal class, and to avoid any kind of confusion, we will henceforth use the following notational conventions: $\xi_u$ and $\tau_u$  denote the localization length and Heisenberg time in the unitary class whereas $\xi_o$ and $\tau_o$ denote the same quantities in the orthogonal class.

\section{CBS/CFS peak for the random QKR}

\subsection{Qualitative discussion}

\begin{figure}
\includegraphics[width=0.7\linewidth]{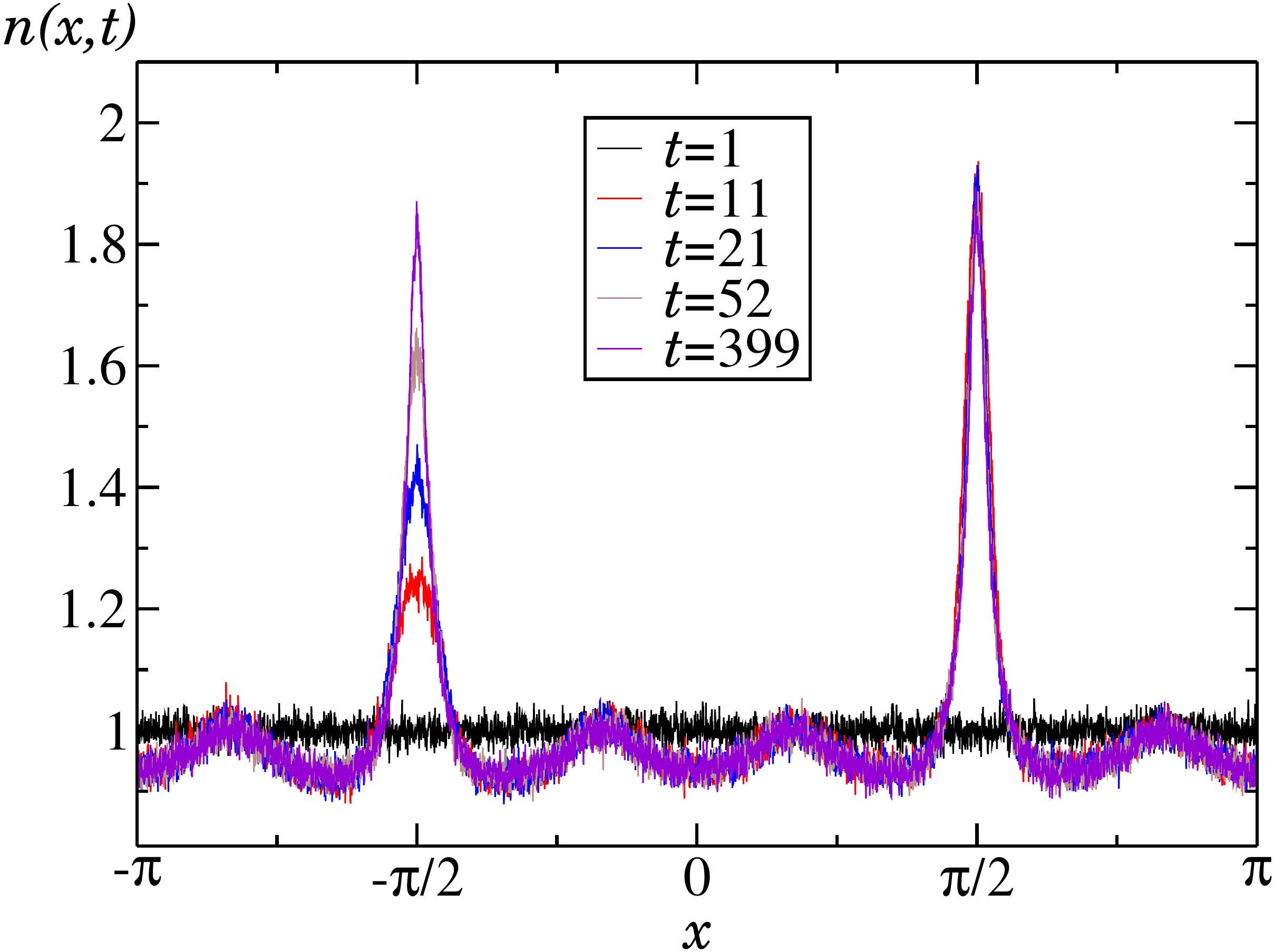}
 \caption{ Time evolution of the real-space density starting from a wave packet initially peaked at $x = x_0=-\pi / 2$ for the random QKR model in the \textit{orthogonal} class described by Eq.~\eqref{eq:HKR}. The parameters are $ K = 20$, $ \hbar_e = 2.89 $ and the Floquet basis size is $ N = 8192$. Just after the first kick, at time $t=1$, the spatial distribution $n(x, t) = \overline{|\langle x|\psi_t\rangle|^2}$ is uniform over $[0,2 \pi]$. At times $t\ge 2$, a CBS peak forms at $x=-x_0 = \pi / 2 $ with contrast close to $1$ while the CFS peak at $x=x_0$ develops after a larger time scale $\tau_o\approx 16$. At times $t\gg \tau_o$ the two peaks are symmetric copies of each other, a direct signature of time-reversal symmetry in the orthogonal class. }
 \label{fig:CBFSdistxK20ortho}
\end{figure} 

If the direct space is defined as the space where Anderson localization occurs, then the CBS/CFS peak structures appear in the associated reciprocal space. For the QKR system, localization occurs in momentum space and the reciprocal space is then the space of positions. From an experimental view, by adiabatically loading an interaction-free and sufficiently cold Bose-Einstein condensate into a deep optical lattice, one can prepare a finite comb of states, each being localized in the wells of the optical lattice. As an idealized version, we consider in the following an infinite initial wave packet $|\psi_0\rangle \sim \sum_{n\in\mathbbm{Z}} |x_0+2\pi n\rangle$ peaked at periodic positions and normalized within each well to $\int_{-\pi}^{\pi} |\psi_0(x)|^2 \, dx/(2\pi) = 1$.
We then numerically compute the state after $t$ kicks, $|\psi_t\rangle = U^t |\psi_0\rangle$, and analyze its averaged spatial distribution on the unit cell $ x \in [- \pi, \pi] $, $n(x, t) = \overline{|\langle x|\psi_t\rangle|^2}$ normalized according to $\int_{-\pi}^{\pi} n(x,t) \, dx/(2\pi) = 1$ (probability conservation). Since the entire model is space-periodic, it is enough to look at the density in a single unit cell. 
To avoid spurious effects related to the quasi-random character of the QKR, we will first consider the random QKR model mentioned at the end of Sec.~\ref{quantumstroboscopics.sec} and then treat the more realistic deterministic model of Eq.~\eqref{eq:HKR}. We further choose sufficiently large $K$ values so that the Heisenberg time is always much larger than the kick period, our unit of time.

\begin{figure}
\includegraphics[width=0.7\linewidth]{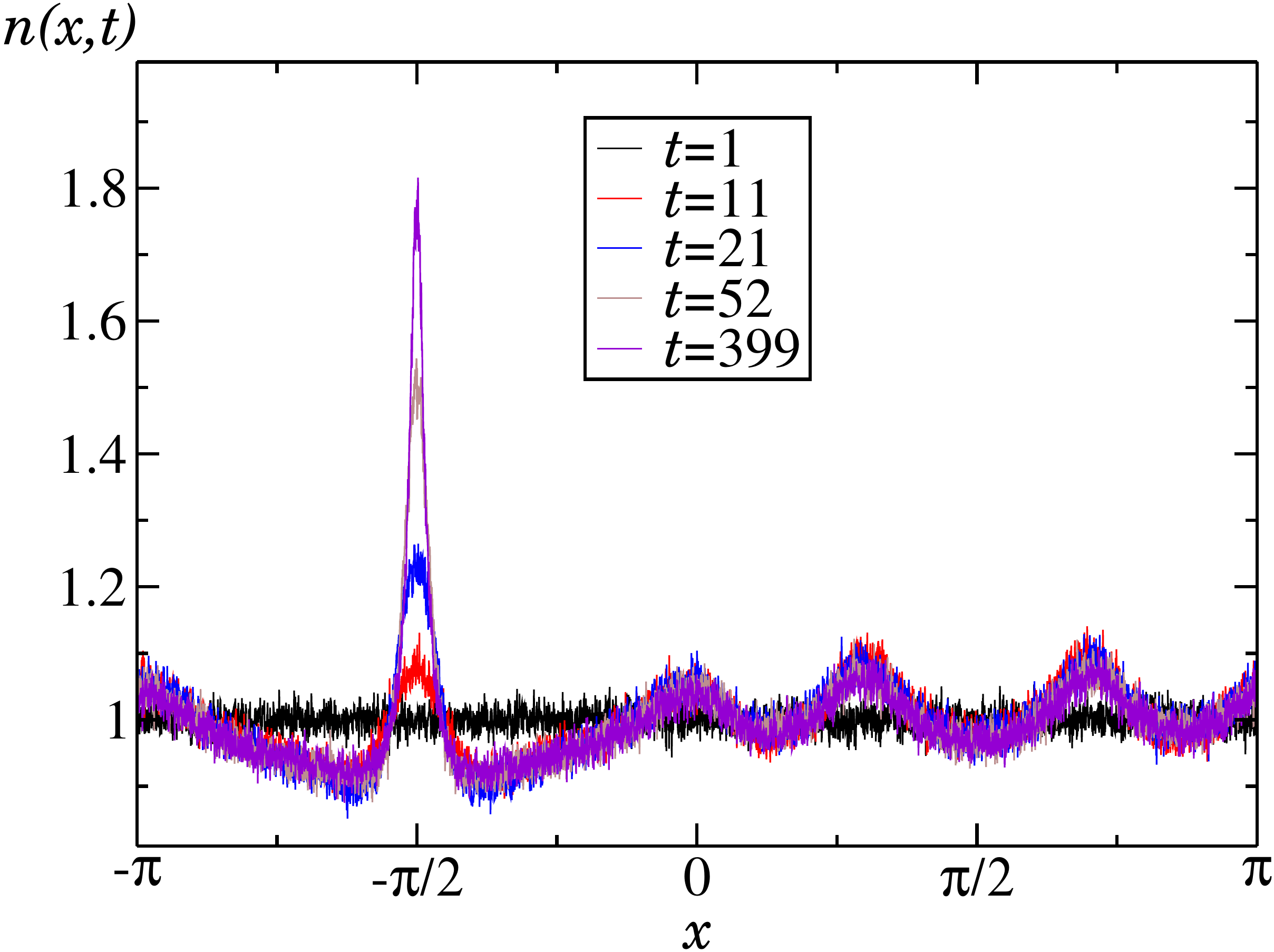}
 \caption{ Same as Fig.~\ref{fig:CBFSdistxK20ortho} but for the random QKR in the \textit{unitary} class described by Eq.~\eqref{eq:kickUC}. Compared to the orthogonal class, the absence of the CBS peak highlights the absence of time reversal symmetry. Nevertheless, the  CFS peak at $x=x_0$ still appears on a time scale $\tau_u\approx 12$. The CFS peak is thus a genuine signature of localization surviving the breaking of time reversal symmetry. The width of the CFS peak is smaller in the unitary class than in the orthogonal class, in accordance with the theoretical prediction Eq.~\eqref{eq:locaQKR-U}.}
 \label{fig:CFSdistxK20uni}
\end{figure}

Figure \ref{fig:CBFSdistxK20ortho} represents the time evolution of our initial wave packet for 
$ x_0 = - \pi/2 $ in the orthogonal class. It is easily seen that the spatial distribution after one kick is uniform in the unit cell, $n(x, t\!\!=\!\!1) = 1$. The CBS peak appears at $x=-x_0 = \pi / 2 $ at the second kick with a contrast already close to $ 1 $. This is easily understood because the scattering mean free time $\tau_s$ identifies here with the kicking period, i.e. $\tau_s = 1 $. As $t$ increases, the CBS contrast reaches 1 very quickly, and its width decreases and saturates at $ \sigma_\text{CBS} = 1 / \xi_o $, the inverse of the localization length, for $ t \gg \tau_o $.
The CFS peak at $ x = x_0$ appears only at longer times $ t \sim \tau_o\gg \tau_s$ and rises 
with a time scale $\sim \tau_o$.  For $ t \gg \tau_o $, it becomes the exact twin of the CBS peak with a width $ \sigma_\text{CFS} = \sigma_\text{CBS} =1/\xi_o$ \cite{lee2014dynamics, ghosh2015cbs3d, ghosh2016cfs3d}.

Figure \ref{fig:CFSdistxK20uni} represents the time evolution of the same initial wave packet in the unitary class, that is subjected to $ \tilde{V}_{\text{kick}}(x)$, Eq.~\eqref{eq:kickUC}. 
As predicted, only the CFS peak is visible at long times $t\gg \tau_u$ while the CBS contrast remains $0$ at all times \cite{micklitz2014strong}. Indeed, while both CBS and CFS effects rely crucially on subtle quantum interference effects, the first one is highly sensitive to time reversal symmetry, but the second one relies only on Anderson localization, a phenomenon surviving the breaking of time reversal symmetry.
These two peak structures are therefore interesting markers of the presence or not of time reversal symmetry in Anderson localization: a system in the unitary class presents only CFS while a system in the orthogonal class presents both CFS and CBS. As a consequence, only CFS is a genuine marker of Anderson localization.

\subsection{Theoretical description}

A detailed theoretical description of the dynamics of the CFS peak for spatial disorder in 1D systems was given in \cite{lee2014dynamics, micklitz2014strong}. Here, using the Floquet eigenbasis, the spatial distribution at time $t$ reads:
\begin{equation}
\label{eq:nt}
n(x,t) = \overline{|\sum_i e^{i\omega_i t} \, \phi^*_i(x_0) \phi_i(x)|^2}
\end{equation}
Since the CBS and CFS peaks are due to interference effects, they should disappear in the presence of dephasing processes, leaving just the so-called incoherent background $n_I(x)$, which is time-independent as soon as $t\gg \tau_s$. We then write $n(x,t) = n_I(x) + n_C(x,t)$ to distinguish the coherent contribution $n_C(x,t)$ from the incoherent one. 

It can be shown \cite{cherroret2012coherent,lee2014dynamics} that the incoherent contribution reads:
\begin{equation}\label{nIx}
 n_I(x)=\int \frac{dE}{2 \pi} \frac{\mathcal{A}(x,E) \; \mathcal{A}(x_0,E)}{2 \pi \nu(E)} \; ,
\end{equation}
where 
$\mathcal{A}(x,E)$ is the disorder-averaged spectral function and $ \nu(E) $ is the disorder-averaged density of states. Since, by definition, $\int \mathcal{A}(x,E) \, dx/(2\pi) = 2\pi \nu(E)$ and $\int \mathcal{A}(x,E) \, dE/(2\pi) = 1$, the incoherent contribution by itself already satisfies probability conservation on the unit cell, $\int n_I(x) \, dx/(2\pi)=1$. 

For matter waves scattered by a spatially random potential, the spectral function 
is far from constant, both as function of energy and momentum (see e.g. \cite{trappe2015semiclassical,prat2016semiclassical}). This induces a non-trivial annular shape of the incoherent background in momentum space \cite{karpiuk2012coherent} that complicates the precise observation of coherence peaks.  
By contrast, one of the main advantages of the QKR is that transport properties are insensitive to energy, and that the background density \eqref{nIx} is flat, $n_I(x) = 1$. Indeed, for the QKR we have 
\begin{eqnarray}
\nu(E) &=& \lim_{N\to\infty} \frac{1}{N} \sum_{i=1}^N  \overline{\delta(E-\omega_i)}
\end{eqnarray}
where $N$ is the Floquet basis size and the relevant integration ranges on $x$ and $E$ are over $2\pi$. 
Since the quasi-energies are uniformly distributed on the circle, $\nu(E)$ is independent of $E$, and from $\int \nu(E) \, dE =1$, one finds $\nu(E) = 1/(2\pi)$. Furthermore, one can show that 
\begin{equation}
\mathcal{A}(x,E) = \lim_{N\to\infty}\frac{2\pi}{N} \sum_{i=1}^N \overline{\delta(E-\omega_i) \, |\phi_i(x)|^2}  
\end{equation}
is also just a constant, $\mathcal{A}(x,E) = 1$. Indeed, its double Fourier transform to momentum and time, $\mathcal{A}(p,t)$, is given by very simple matrix elements of $U^t$ averaged over disorder: 
\begin{equation}
\label{eq:KroDel}
\overline{\langle p | U^t | p'\rangle} = 2\pi \delta(p-p') \, \delta_{0t}.
\end{equation}
As a consequence, the incoherent background of Eq.~\eqref{nIx} is spatially uniform, $n_I(x) = 1$, as we have duly checked numerically (results not shown here). 

Our figure of merit then is the contrast $C_{F/B}(t) = n_C(\pm x_0,t)/n_I$ of the coherence peaks at time $t$, i.e. the height of the coherent contribution relative to the background, evaluated at $+x_0$ for the CFS peak and at $-x_0$ for the CBS peak. Since both the total density $n(x,t)$ and the background are normalized to 1 on each unit cell, the coherent contibution must satisfy $\int n_C(x,t) \, dx =0$. As a consequence, the coherent contribution $n_C(x,t)$ cannot feature only positive peaks, but must as well take negative values somewhere in the unit cell.  

The CBS/CFS peak contrasts are then
\begin{eqnarray}\label{eq:defCBS}
 C_B(t)&=&n(-x_0,t) -1  \; , \\
 \label{eq:defCFS}
 C_F(t)&=& n(x_0,t) -1 \; .
\end{eqnarray}
At very long times $t\gg \tau$, $n_t(x)$ reaches the stationary spatial distribution
\begin{equation}
n_S(x) = \lim_{N\to\infty} \sum_{i=1}^{N} \overline{\vert \phi_i(x) \vert ^2 \vert \phi_i(x_0) \vert ^2 } 
\end{equation}
given by the diagonal term in Eq.~\eqref{eq:nt}. The converged, stationary contrasts are thus:
\begin{eqnarray}
 C^\infty_B &=& n_S(-x_0) -1 \\ 
 C^\infty_F &=& n_S(x_0) -1.
\end{eqnarray}
As shown in \cite{lee2014dynamics} for correlated spatial disorder, one has $C^\infty_F=1$ when the localization length is much larger than $\ell_s$ but $C^\infty_F< 1$ when the two length scales are comparable. For the QKR, we thus get $C^\infty_F=1$ for $K\gg \hbar_e$. Furthermore, for systems belonging to the orthogonal class, it can be shown that time reversal invariance enforces $ C^\infty_B =  C^\infty_F$. In the orthogonal class, the CBS and CFS peaks are then mirror images of each other in the long-time limit. In the unitary class, the CBS peak is absent while the CFS peak persists.


\subsection{Scaling behavior}

\subsubsection{Unitary Class}

\begin{figure}
\includegraphics[width=0.7\linewidth]{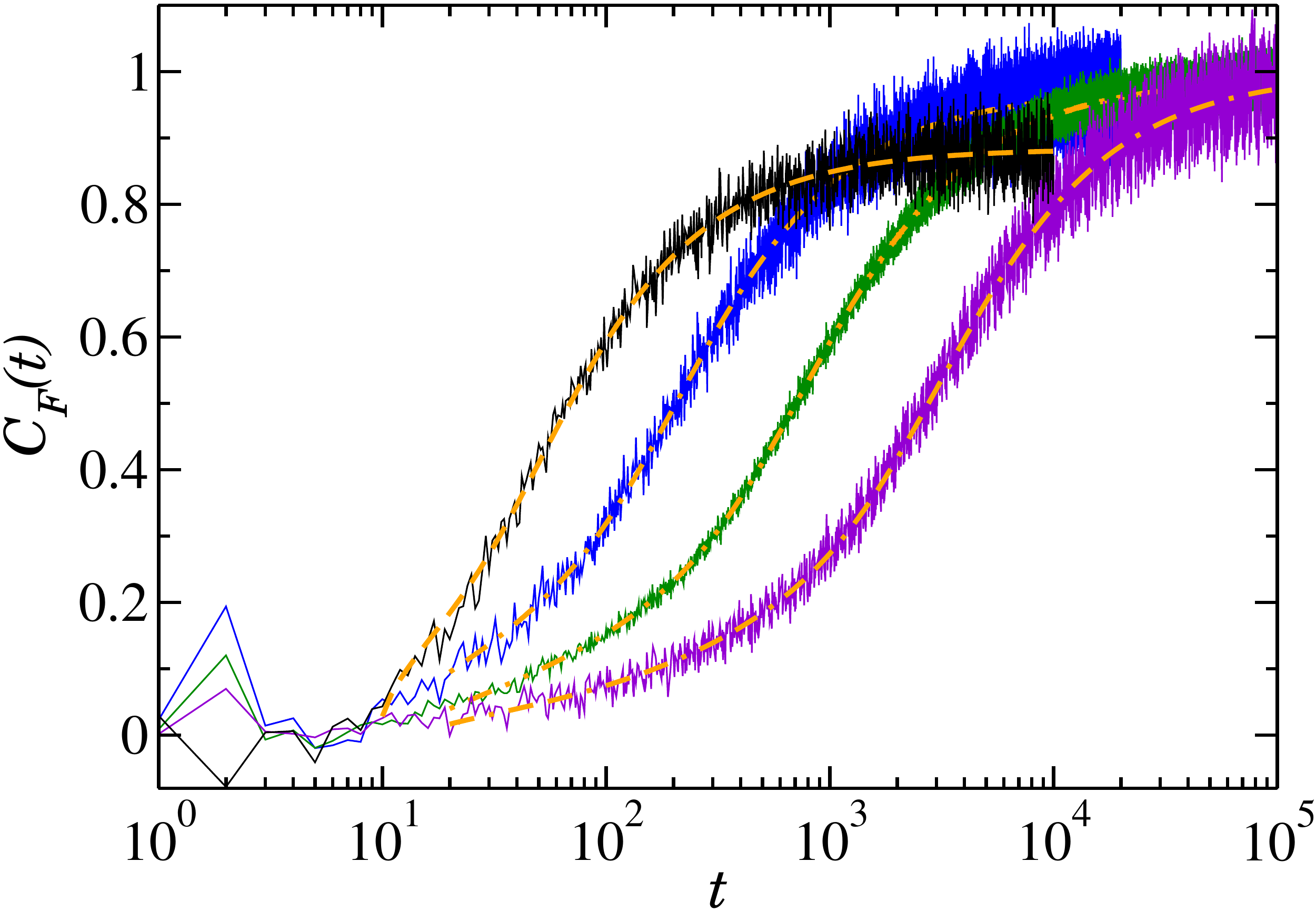}\\
\includegraphics[width=0.7\linewidth]{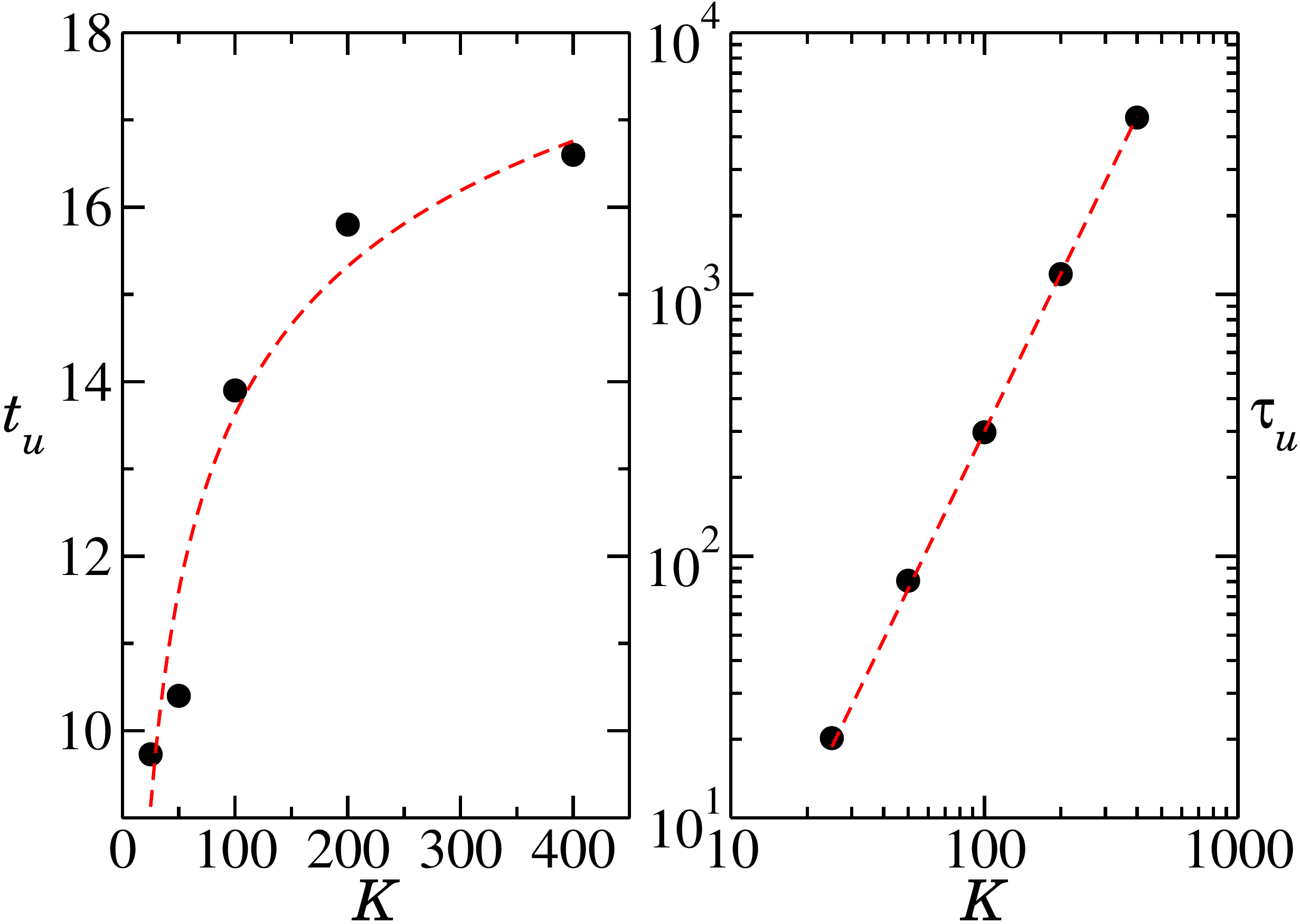}
 \caption{Dynamics of the CFS peak in the \textit{unitary} class.
Upper panel: Temporal evolution of the CFS contrast Eq.~\eqref{eq:defCFS} for the random QKR run with Eq.~\eqref{eq:kickUC} and $\hbar_e=2.89$. Numerical curves from left to right: $K=25$, $50$, $100$ and $200$. The Floquet basis size $N$ varies from $8192$ to $32768$. At long times $ t \gg \tau_u$, $C_F(t)$ tends to a constant value $ C^\infty_F $ which gets closer to 1 as 
$K$ gets larger. 
Using this stationary contrast together with the Heisenberg time $\tau_u$ and the delay time $t_u$ as fitting parameters, perfect agreement is found with the theoretical prediction Eq.~\eqref{eq:CFStheodelay} (dashed orange lines) over several orders of magnitude. Lower panel: Plot of the fit parameters $t_u$ and $\tau_u$ as a function of $K$.
Left lower panel: The behavior of $t_u$ is well fitted by $t_u=a_1 \times \ln (a_2K/\hbar_e)/\ln (a_3K)$ with $a_1=34$, $a_2=0.62$ and $a_3=22$ (red dashed line), a logarithmic behavior reminiscent of an Ehrenfest time $t_E$, Eq.~\eqref{eq:tEh}. Right lower panel: The behavior of the extracted Heisenberg time $\tau_u$ follows closely the scaling relation Eq.~\eqref{eq:Heisenberg-U} (red dashed line).
 }
 \label{fig:CfsvstUNI}
\end{figure} 

\begin{figure}
\includegraphics[width=0.7\linewidth]{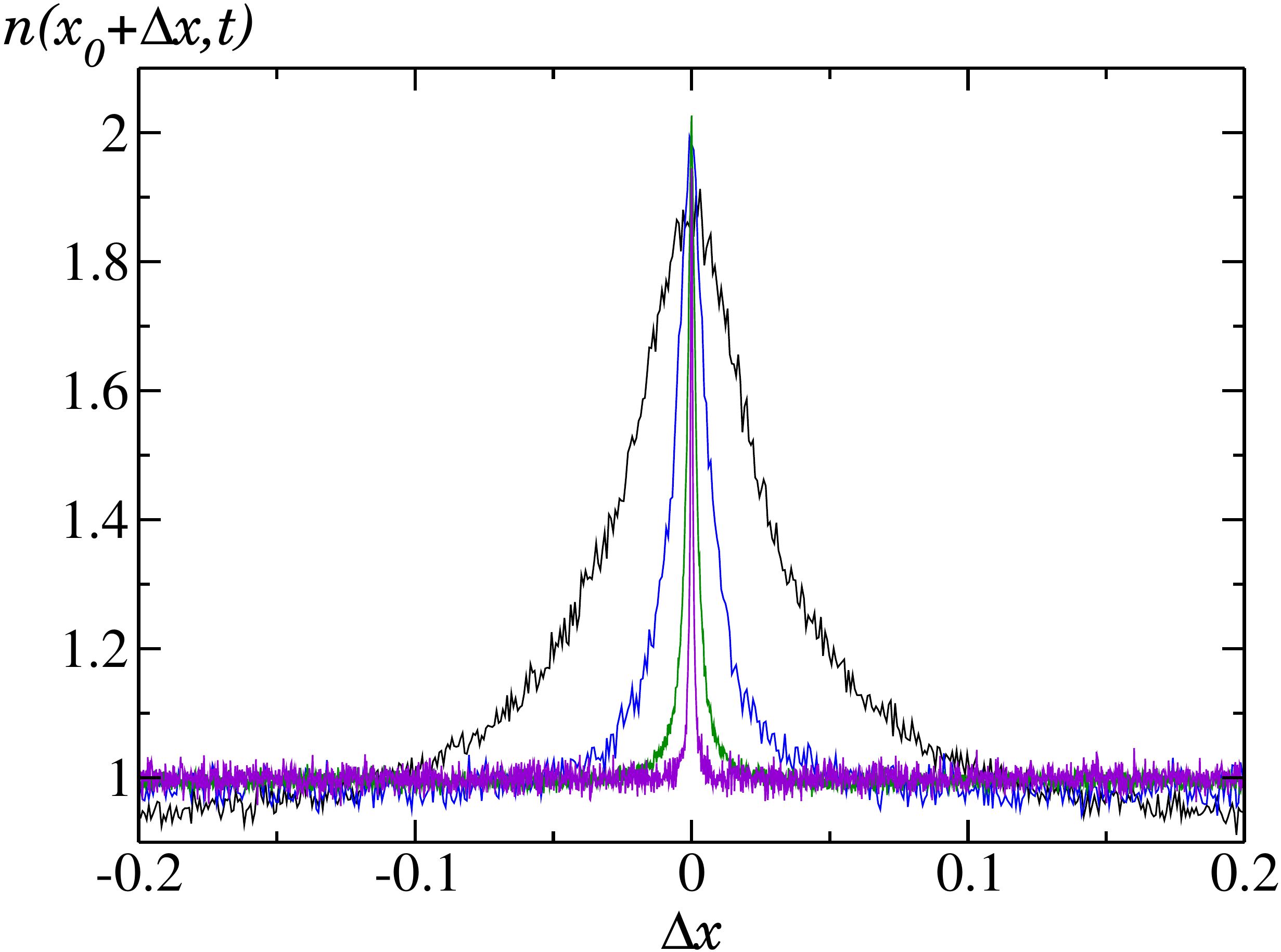}\\
\includegraphics[width=0.7\linewidth]{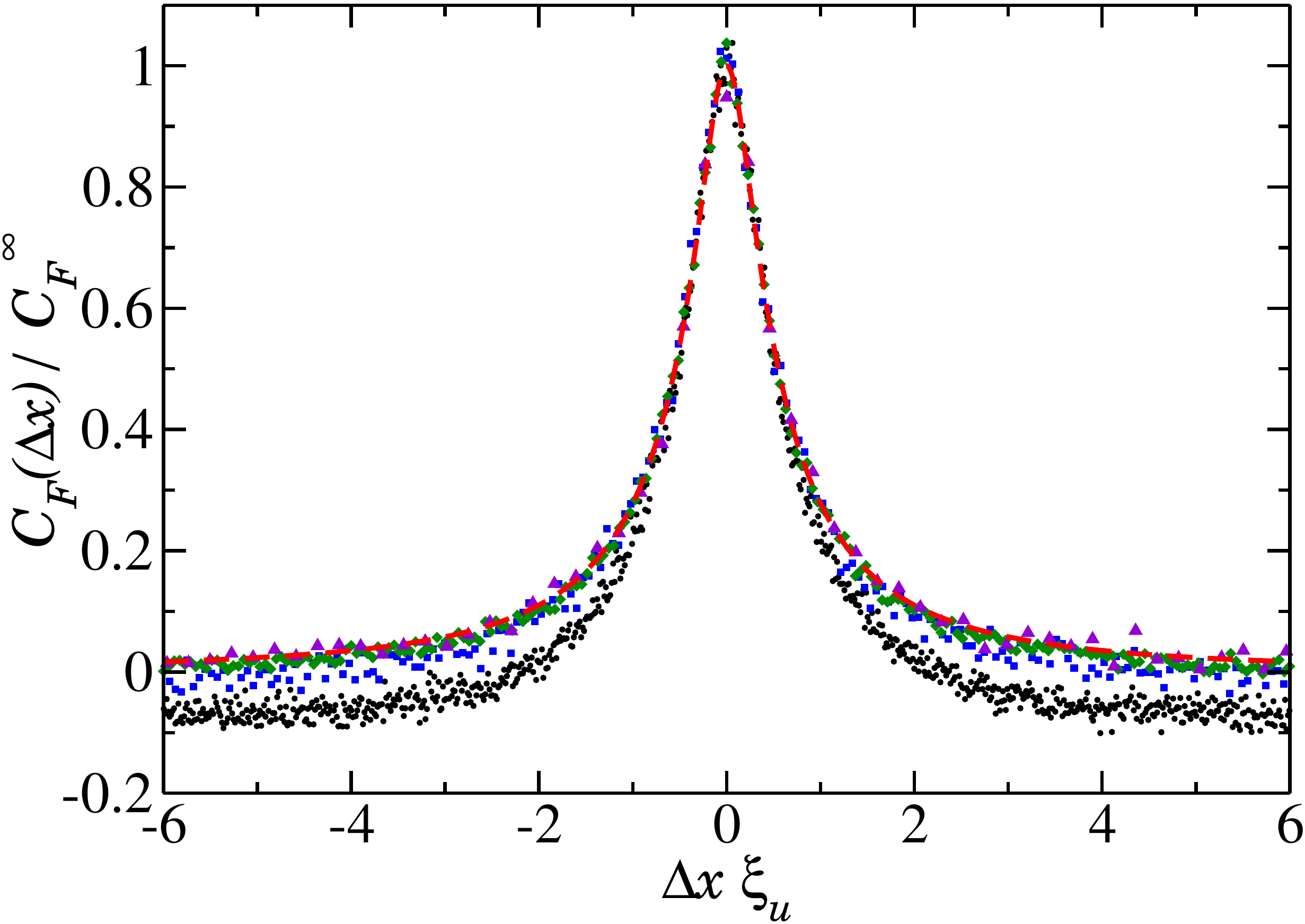}
 \caption{Spatial scaling behavior of the CFS peak at infinite time in the \textit{unitary} class. Upper panel: Averaged spatial distribution $n(x_0+\Delta x,t)$ around the initial position $x_0$ obtained at a time $t\gg \tau_u$ for the random QKR run with Eq.~\eqref{eq:kickUC} and $\hbar_e=2.89$.  From broader to narrower peaks: $K=25$, $t=10^4$; $K=50$, $t=2\times 10^4$; $K=100$, $t=10^5$; $K=200$, $t=10^5$. The Floquet basis size $N$ ranges from $8192$ to $32768$. Lower panel: Spatial scaling behavior of the CFS contrast at long times around $x_0$: $K=25\, (\text{black }\circ),\, 50\, (\text{blue }\square),\, 100\, (\text{green }\diamond)$ and $200\, (\text{violet }\triangle)$. 
When $\Delta x$ is rescaled by $1/\xi_u$ with $\xi_u=\tau_u = K^2/(4\hbar^2_e)$, the CFS peak shapes collapse beautifully onto a single scaling curve. The dashed red line is the function $\mathcal C_1$ in Eq.~\eqref{eq:Cfsvsxtheo} and taken from \cite{micklitz2014strong} (no fitting parameter). Very good agreement is found at large $K$ when negative wings of the CFS peak, required by probability conservation, are negligible.
}
 \label{fig:CfsvsxUNI}
\end{figure} 


The time dependence of the CFS contrast in the unitary class has been predicted in Ref.~\cite{micklitz2014strong} to be of the form $C_F(t) = C^\infty_F \, F(t)$, with 
 \begin{equation}\label{eq:CFStheo}
 F(t)= I_0(2 \tau_u/t) \; e^{-2\tau_u/t} \; ,
\end{equation}
where $I_0$ is the modified Bessel function of order zero \cite{abrastegun1972} and where $\tau_u$ is the Heisenberg time in the unitary class. 
Using the kicking potential Eq.~\eqref{eq:kickUC}, we have performed the first numerical check of this prediction for the unitary class. Figure \ref{fig:CfsvstUNI} shows the numerical results obtained for the temporal evolution of the CFS contrast in the unitary class and its comparison to Eq.~\eqref{eq:CFStheo}. Our results 
are in perfect agreement with this theoretical prediction {\it provided} the onset is shifted by a certain delay time $t_u$ depending on $K$ and $\hbar_e$:
\begin{equation}\label{eq:CFStheodelay}
 C_F(t)= C^\infty_F \; F(t-t_u) \; .
\end{equation}
Indeed, using $C^\infty_F$, $\tau_u$ and $ t_u$ as fit parameters, Eqs.~\eqref{eq:CFStheo} and \eqref{eq:CFStheodelay}, are able to reproduce accurately the temporal dynamics of the CFS from early times $t=t_u$ to large times $t\gg \tau_u$. The physical origin of this delay time $t_u$ might be related to a Ehrenfest time effect. Indeed, it has been argued in \cite{tian2004weak} that dynamical localization should be delayed by a time scale given by the Ehrenfest time $t_E$ of the system:
\begin{equation}\label{eq:tEh}
t_E = \frac{\ln(K/\hbar_e)}{2\ln(K/2)}
\end{equation}
In Figure \ref{fig:CfsvstUNI}, we have plotted the numerically-found delay times $t_u$ as a function of $K$ at $\hbar_e=2.89$ and compared them to the fit function $a_1 \times \ln (a_2K/\hbar_e)/\ln (a_3K)$. The agreement is reasonable and tends to support a logarithmic dependence of $t_u$ on the stochasticity parameter. We have also plotted $\tau_u$ against $K$ and found a perfect agreement with the scaling relation:
\begin{equation}
\label{eq:Heisenberg-U}
\tau_u = \frac{K^2}{4\hbar^2_e}
\end{equation}

The theory developed in \cite{micklitz2014strong} also predicts the shape of the CFS peak at infinite times. For the QKR, it reads
\begin{equation}\label{eq:Cfsvsxtheo}
C_F(\Delta x)\equiv \lim_{t \rightarrow \infty} n(x_0+\Delta x,t) -1 = C^\infty_F \; \mathcal C_1(\Delta x \, \xi_u);
\end{equation}
the function $\mathcal C_1$ is defined by an integral whose analytical expression can be looked up in Ref.~\cite{micklitz2014strong}. We have tested the accuracy of this prediction for the peak shape by using the values previously found for $ C^\infty_F $. The results are shown in Figure \ref{fig:CfsvsxUNI}, with good agreement provided we choose $ \xi_u = \tau_u$ (see Eq. \eqref{eq:locaQKR}). Obvious deviations can be ascribed to negative wings of the CFS contribution that are not accounted for by the theoretical prediction \cite{micklitz2014strong}. We remind the reader that these negative wings are necessary to ensure normalization: they cancel the positive peak contribution around $x_0$. Since this positive excess is concentrated over the width $1/\xi_u$, the negative wings get shallower when $\xi_u$ increases with increasing $K$, which is indeed what we observe. 


\subsubsection{Orthogonal Class}

\begin{figure}
\includegraphics[width=0.7\linewidth]{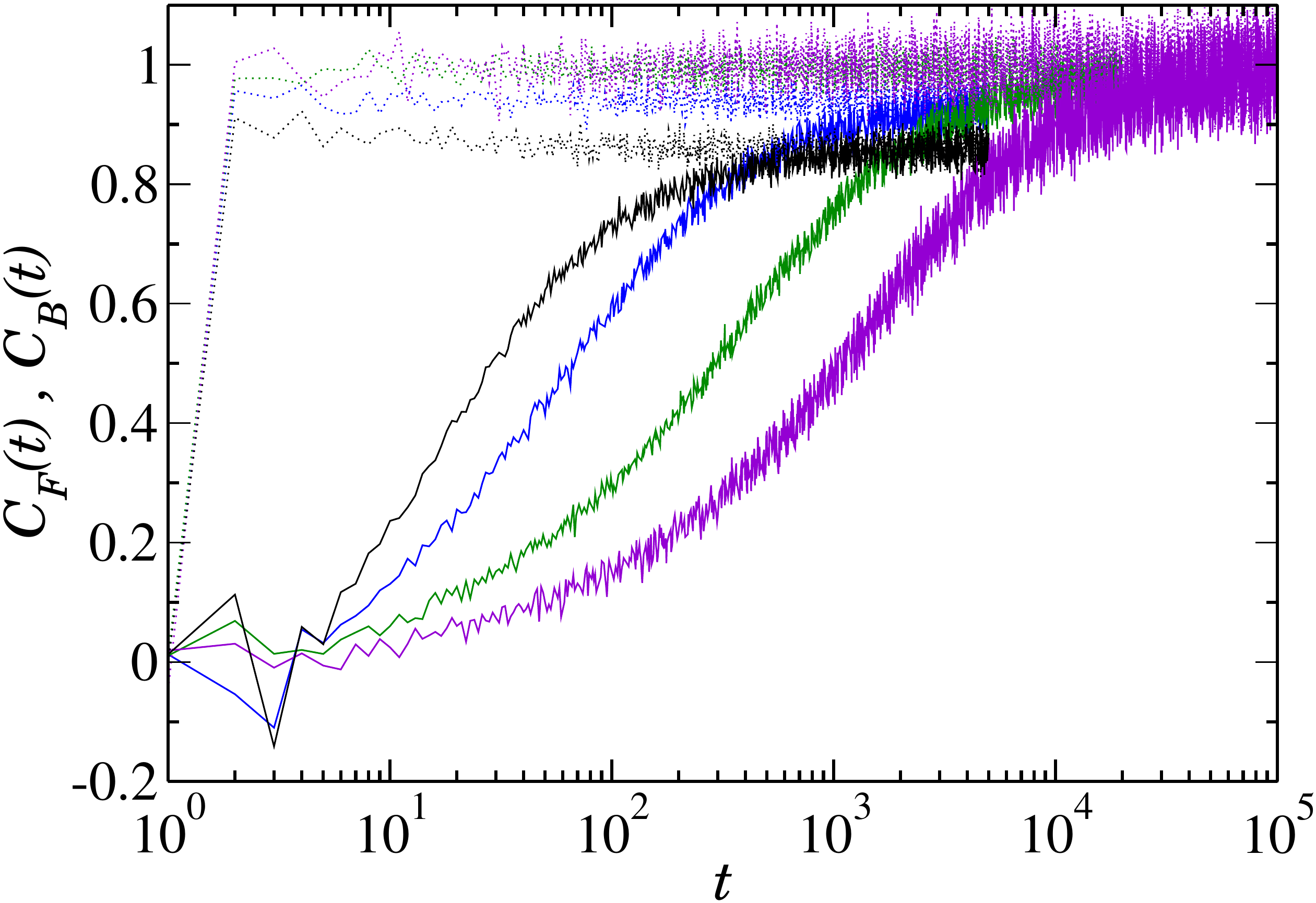}\\
\includegraphics[width=0.7\linewidth]{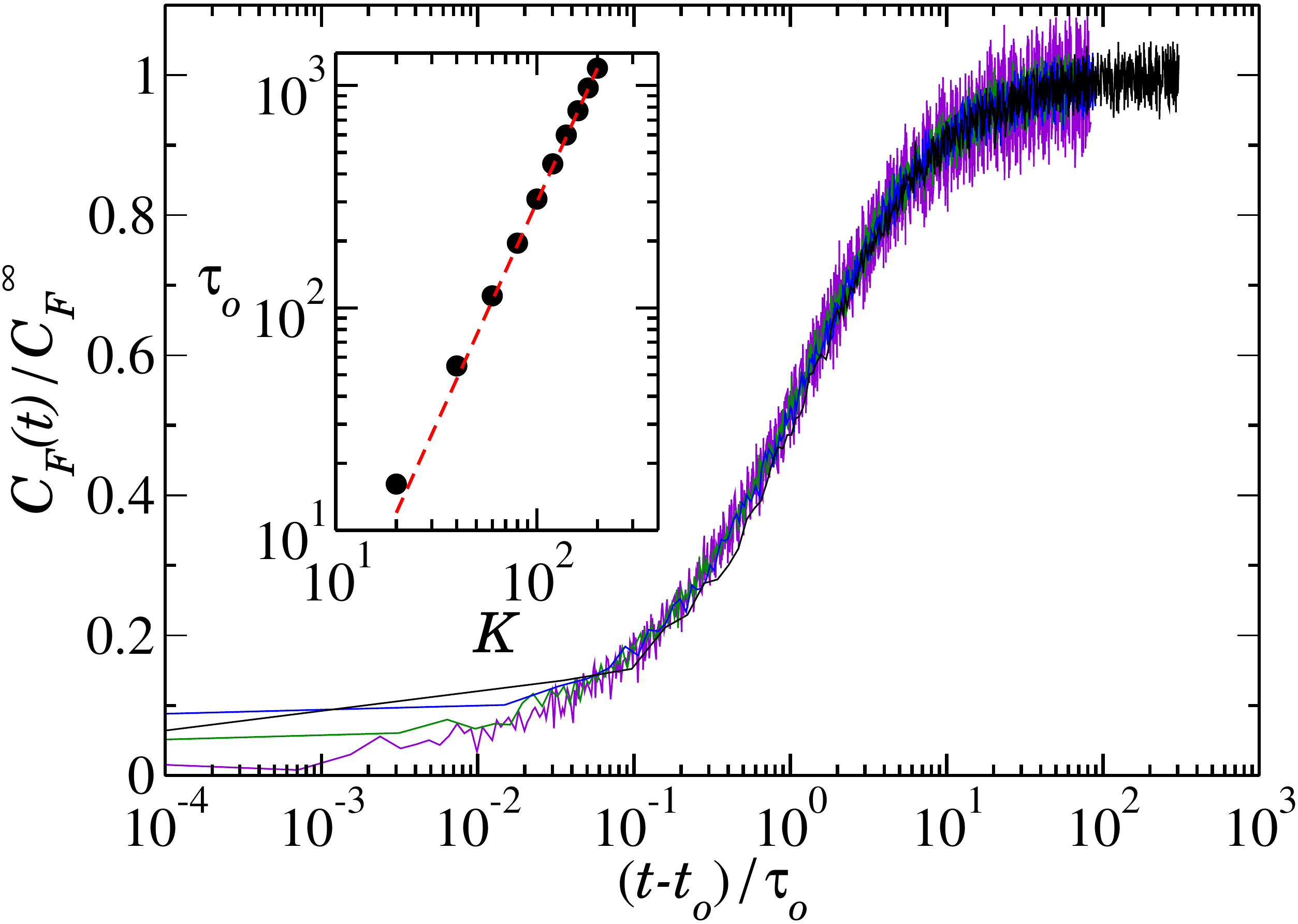}
 \caption{Dynamics of the CBS/CFS contrast in the \textit{orthogonal} class. Upper panel:
Unscaled temporal evolution of the CBS/CFS contrast (dashed/full lines). From left to right: $K=20$, $40$, $100$ and $200$. The Floquet basis size $N$ ranges from $8192$ to $32768$, and $\hbar_e=2.89$. $C_F(t)$ tends to $C_B(t)$ at long times $ t \gg \tau_o$, with $\tau_o$ the Heisenberg time in the orthogonal class. 
Lower panel: Scaling behavior of $C_F(t)$ following Eq.~\eqref{eq:CFStheodelay} with two new characteristic times, $\tau_o$ and $t_o=2 t_u/3$ (see text). The contrasts collapse for different $K$ values to a universal curve that 
deviates slightly, but systematically from the theoretical prediction Eq.~\eqref{eq:CFStheo} for the unitary class (not shown). 
The Heisenberg time $\tau_o$ has been determined independently from the dynamics of a wave packet in momentum space. The inset shows that it compares well with the theoretical prediction Eq.~$\eqref{eq:Heisenberg-U}$ (red dashed line). 
 }
 \label{fig:CfbsvstORTHO}
\end{figure}

\begin{figure}
\includegraphics[width=0.7\linewidth]{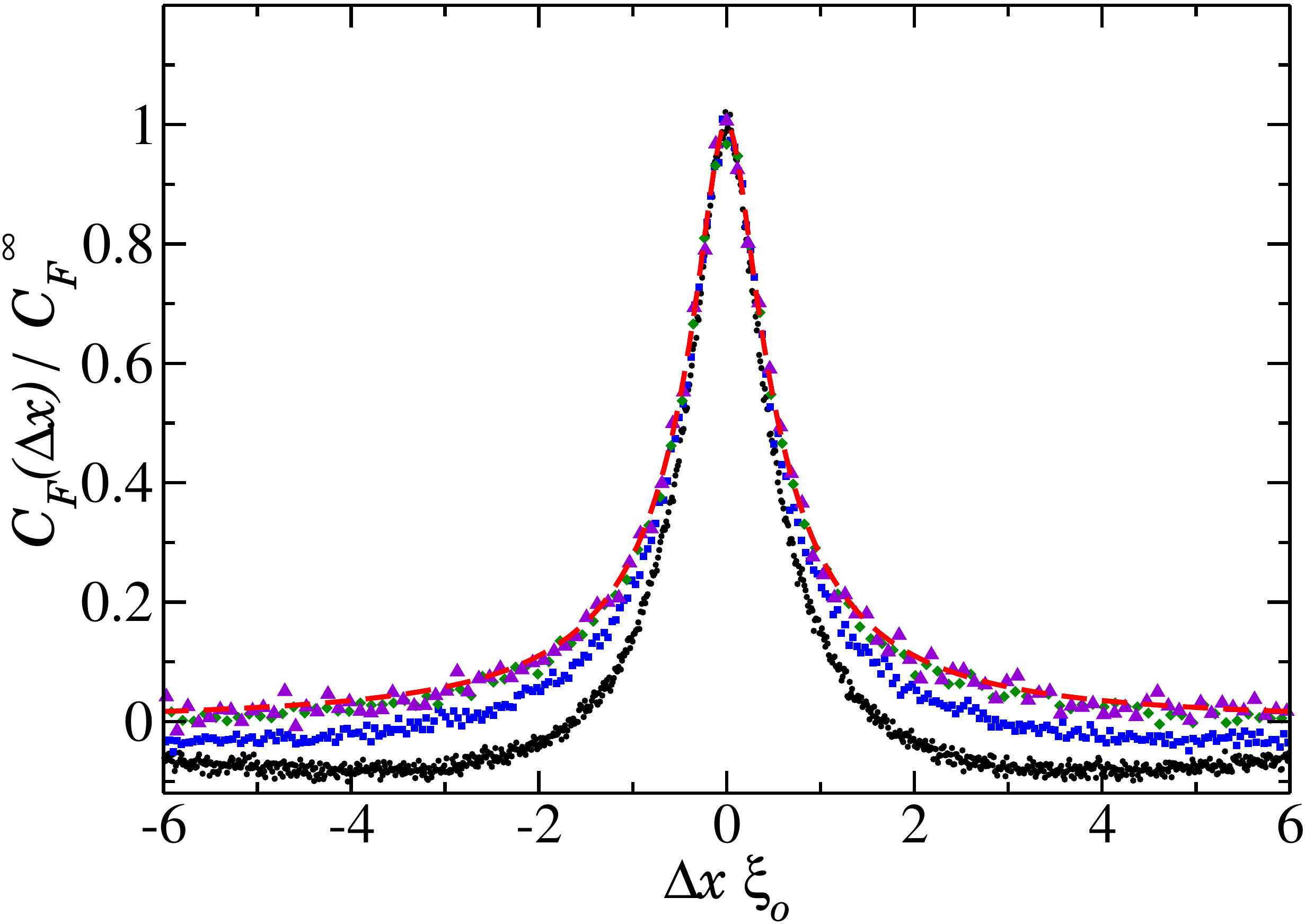}
 \caption{Scaling spatial behavior of the CFS peak at infinite time in the \textit{orthogonal} class. The distance $\Delta x$ to the initial position $x_0$ is here rescaled by $1/\xi_o$, where $\xi_o=\tau_o/2$ is the localization length in the orthogonal class (see text). The CFS peaks for the different values $K=20\, (\text{black }\circ),\, 40\, (\text{blue }\square),\, 100\, (\text{green }\diamond)$ and $200\, (\text{violet }\triangle)$ (with $\hbar_e = 2.89$) collapse onto a single scaling curve. The red dashed line shows the theoretical prediction obtained directly  from the unitary result Eq.~\eqref{eq:Cfsvsxtheo} by just replacing $\xi_u$ by $\xi_o$. Surprisingly good agreement is found at large $K$ values where the probability-conserving wing contributions are negligible.
 }
 \label{fig:CfsvsxORTHO}
\end{figure} 

Figure~\ref{fig:CfbsvstORTHO} shows the temporal evolution of the CBS and CFS contrasts, Eqs~\eqref{eq:defCBS}-\eqref{eq:defCFS}, in the \textit{orthogonal} class. While $C_B(t)$ is already a constant $C_B^\infty$ (close to $1$ for $K$ large enough) at $t\geq2$, $C_F(t)$ reaches $C_F^\infty=C_B^\infty$ only at large times $t \gg \tau_o$ with $\tau_o$ the Heisenberg time in the orthogonal class. 
To the best of our knowledge there is no theoretical prediction for the CFS contrast and peak shape in the orthogonal class. Despite a behavior similar to the unitary case, the predictions Eqs.~\eqref{eq:CFStheo}-\eqref{eq:CFStheodelay} fail to accurately reproduce the observed time dependence of the contrasts, see \cite{lee2014dynamics} for a similar observation in systems with spatial disorder. 
Nevertheless, as shown in the lower panel of Fig.~\ref{fig:CfbsvstORTHO}, all curves still collapse onto a single curve if we use a scaling similar to the unitary class and plot the contrasts against $(t-t_o)/\tau_o$ where $t_o$ is the delay time in the orthogonal class. Following the analogy between $t_u$ and the Ehrenfest time $t_E$ in the unitary class, and accounting for the known Ehrenfest time effects on weak localization discussed in \cite{tian2004weak}, we set $t_o=2 t_u /3$ in the orthogonal class (compare Eqs.~(5) and (102) in \cite{tian2004weak}). Moreover, we have independently determined $\tau_o$ from the dynamics of a wave packet in momentum space and found good agreement with the relation $\tau_o \approx \tau_u = K^2/(4\hbar^2_e)$ when $K$ is large enough (see inset in the lower panel of Fig.~\ref{fig:CfbsvstORTHO}). The lower panel of Fig.~\ref{fig:CfbsvstORTHO} shows a good collapse of the curves for different values of $K$ over a large range of times. On the contrary, not taking into account $t_o$ results in significant deviations at short times (data not shown). This confirms the importance of the delay time and its analogy with the Ehrenfest time.

Finally, the CFS peak shape at large times $t\gg \tau_o$ is shown in Fig.~\ref{fig:CfsvsxORTHO}. Again, a scaling behavior is observed when the distance $\Delta x$  to the initial position $x_0$ is rescaled by $1/\xi_o$, where $\xi_o=\tau_o/2$ is the localization length in the orthogonal class. The factor $1/2$ in the definition of $\xi_o$ can be understood from Eq.~\eqref{eq:locaQKR-U} and from the previously found relations $\tau_o \approx \tau_u = \xi_u = K^2/(4\hbar^2_e)$.
Quite surprizingly, the function $ \mathcal{C}_1$  found for the unitary class, Eq.~\eqref{eq:Cfsvsxtheo} from \cite{micklitz2014strong}, describes accurately the spatial dependence of the CFS peak in the orthogonal case. Deviations are again observed when $K$ is not large enough due to negative wings in the CFS peak, arising from probability conservation constraints.

\section{Proposed experimental protocol to observe the CFS and CBS peaks}

Even though the CBS and CFS properties do not depend on energy, a direct experimental observation of these peak structures with the QKR is difficult because it requires to measure atomic densities on spatial scales significantly smaller than the lattice constant of the kick potential. One can however bypass this bottleneck by performing a phase space rotation (as explained below), converting hereby the spatial structures of the CBS and CFS peaks into well-defined, and more easily measurable signatures in the momentum distribution. Such a method has been recently experimentally demonstrated \cite{ourPRL2016}.

\begin{figure}
\includegraphics[width=0.7\linewidth]{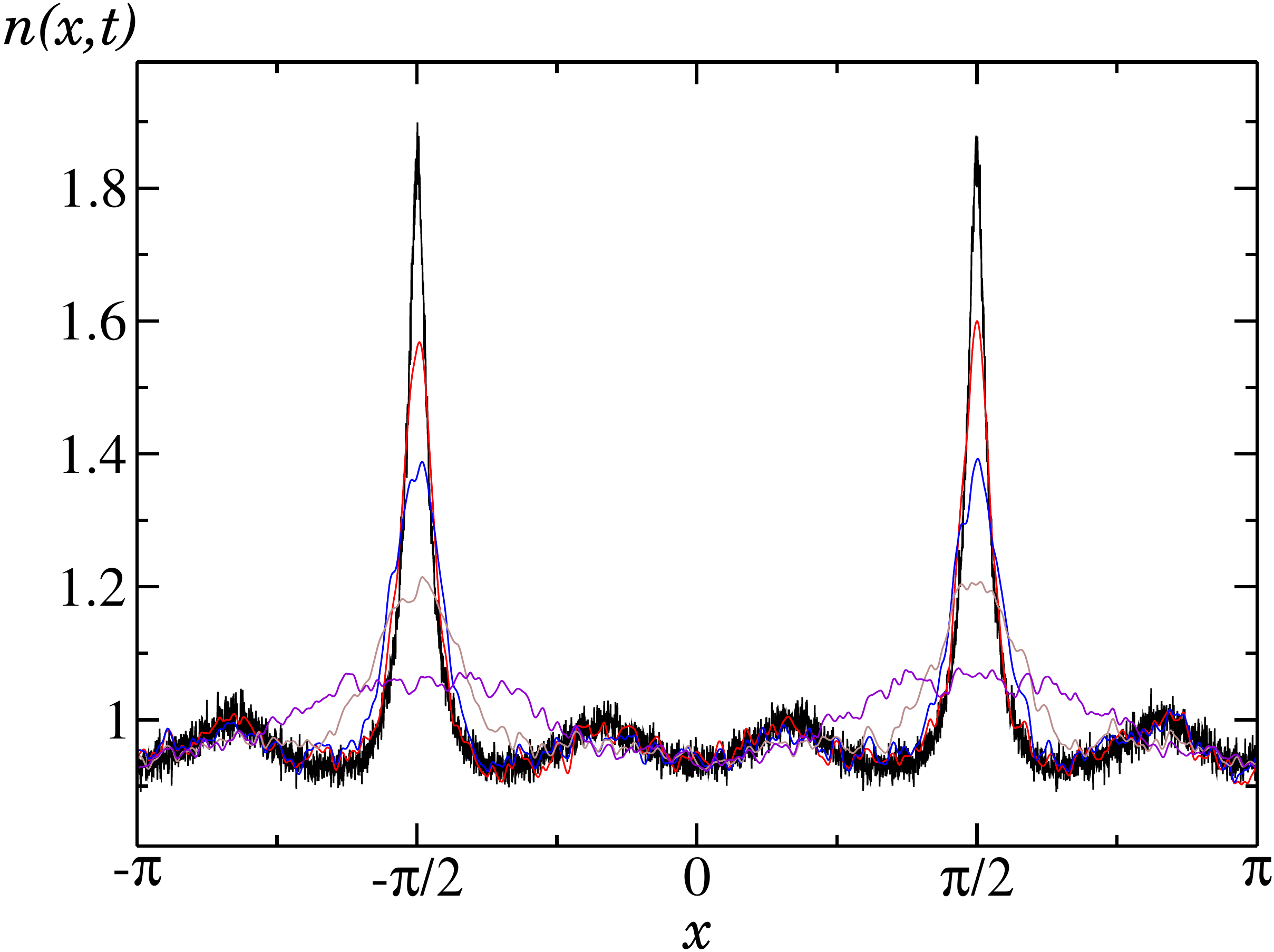}\\
\includegraphics[width=0.7\linewidth]{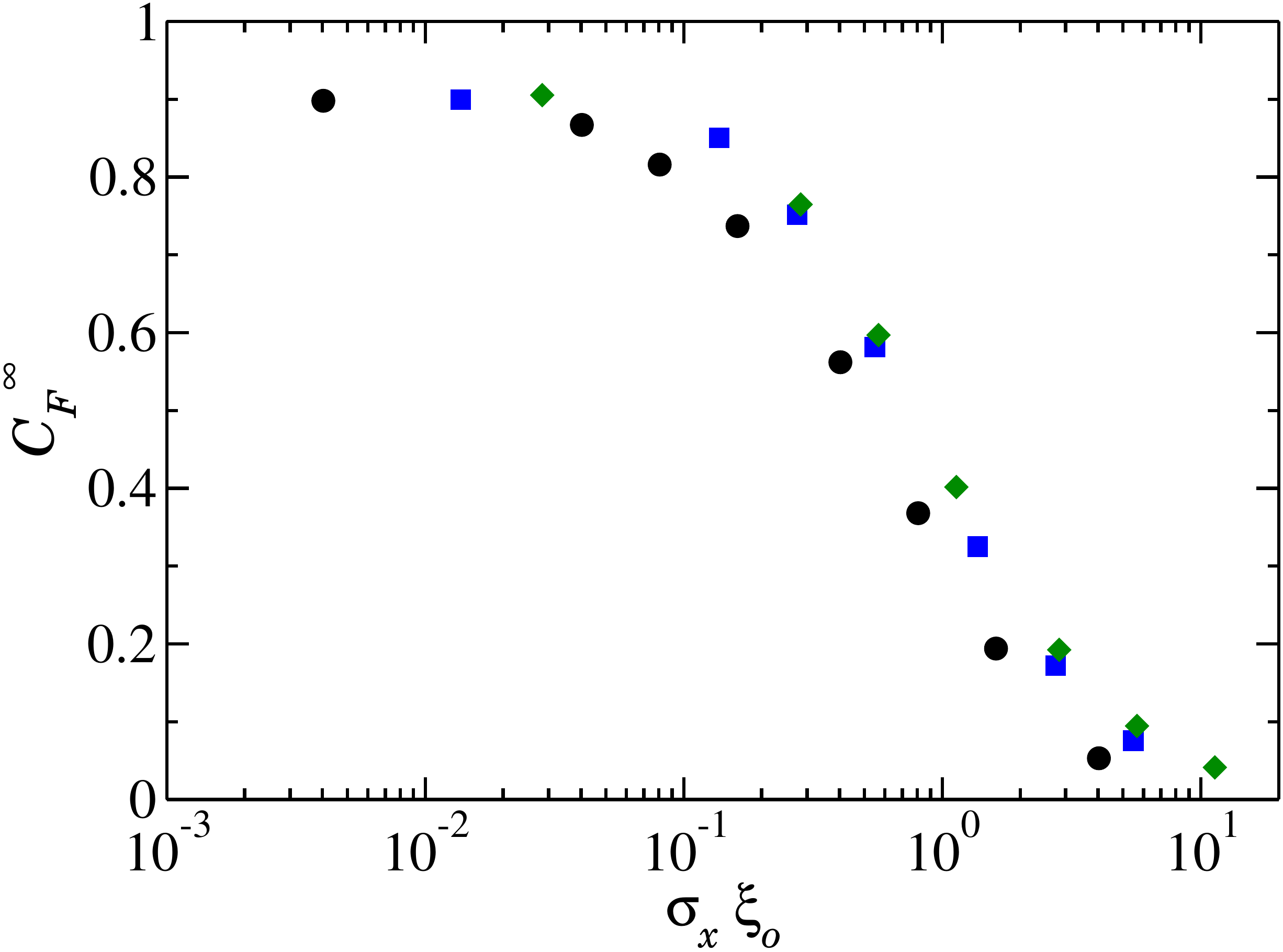}
 \caption{Convolution effect of the spatial width of an initial state on the CBS and CFS peaks structures of the random QKR in the \textit{orthogonal} class. The initial state is a periodic comb of Gaussian wave packets with spatial width $\sigma_x$ and centered at $x_0=-\pi/2$ mod$[2 \pi]$. Upper panel: spatial distribution $n(x,t)$ obtained at large time $t=400\gg \tau_o\approx 16$ for different values of $\sigma_x$ at $K=20$ and $\hbar_e=2.89$ ($\xi_o \approx 8$). From narrower to broader peaks: $\sigma_x=5\times 10^{-4}$, $0.05$, $ 0.1$, $ 0.2$, and $0.5$. As $\sigma_x$ increases, the CBS and CFS peaks are smoothed out by the convolution with the initial state. Lower panel: scaling behavior of $C_F^\infty$ as a function of $\sigma_x\xi_o$ for $K=20\, (\text{black }\circ),\, 40\, (\text{blue }\square),\, 60\, (\text{green }\diamond)$. The contrast is close to one when $\sigma_x \xi_o \ll 1$ whereas it vanishes for $\sigma_x \xi_o \gg 1$.  
 }
 \label{fig:inista}
\end{figure}

\subsection{Initial state} 

An initial state realistically prepared in an experiment will consist of a non-interacting BEC loaded adiabatically into a deep optical lattice achieving $s= 2 V_0/E_R \gg 1$. Actual values of $s$ depend on the atomic species and on the power and waist of the laser beams used to create the optical lattice. Typically, values between $s=10$ and $s=50$ can be readily achieved.  For a deep optical lattice, the initial state is essentially the coherent superposition of the groundstates of the local harmonic wells, {\it i.e.}, a periodic comb of minimal Gaussian wave packets $ \psi(X)=e^{-X^2/2a_0^2}/(\pi a_0^2)^{1/4}$ where $a_0=\sqrt{2}s^{-1/4}/k$ is the local harmonic length. Using the dimensionless units imposed by the lattice, we write the spatial distribution in each well as $ n(x,t=0)=e^{-x^2/2{\sigma_x}^2}/(\sigma_x \sqrt{2 \pi})$. Direct inspection shows that $\sigma_x=s^{-1/4}$, so the deeper the lattice, the better localized the atoms.

In Fig.~\ref{fig:inista} we plot the average density $n(x,t)$ at large times for this initial state as a function of the initial width $\sigma_x$. At long times, we observe a convolution of the CBS/CFS peaks, of size $1/\xi_o$, by the initial state of spatial width $\sigma_x$. Large contrasts, close to $1$, are only observed when $\sigma_x \xi_o \ll 1$ whereas for $\sigma_x \xi_o \gg 1$ the peak structures are smoothed out (see also \cite{micklitz2014strong}). Experimentally, we thus need to have at least $\sigma_x \xi_o < 1$ for a clear observation of the CBS and CFS peaks, knowing that it is difficult to go beyond $s=80$ and thus below $\sigma_x \approx 0.33$ (see below).

\subsection{CBS/CFS peaks for the atomic kicked rotor}

\begin{figure}
\includegraphics[width=0.7\linewidth]{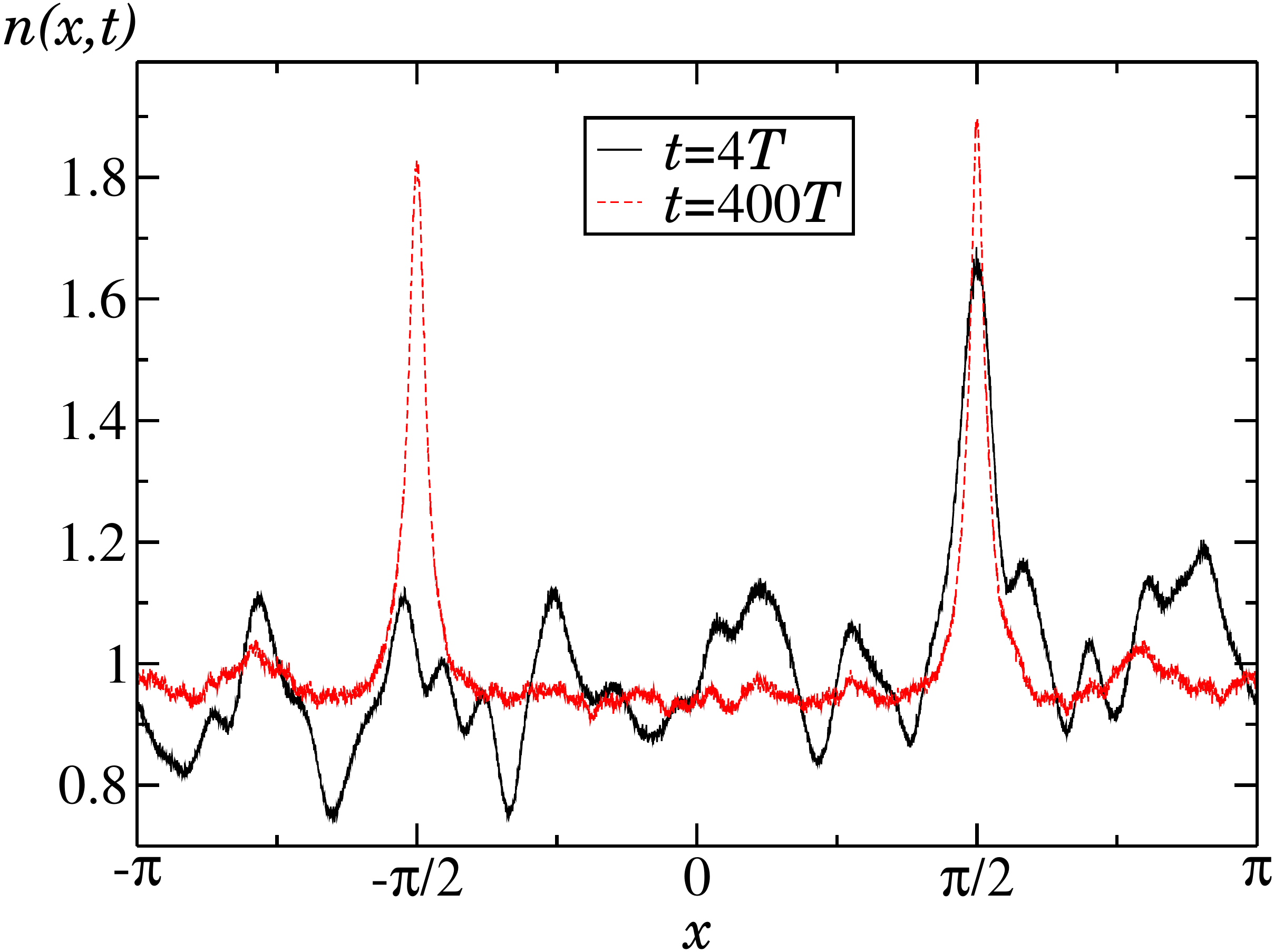}
 \caption{Time evolution of the spatial distribution of a wave packet initially peaked at $ x_0 = - \pi / 2 $ subjected to the modulated KR, Eq.~\eqref{eq:HKRmod}. The plot is obtained for experimentally accessible parameters ($ \omega_2 = 2 \pi \frac{64}{25} $, $ \hbar_e =  2.89$, $K = 6$ and $ \varepsilon =  0.8$) but for an unrealistic vanishing initial width $\sigma_x \ll 1/\xi_o$. The modulation period is $ T = 25 $ kicks. With this choice of parameters, the Heisenberg time $\tau_o$ of the system is of the same order as $T$. At short times, $ t=100 = 4 T $, one observes
the formation of a CBS peak at  $x=-x_0 = \pi / 2 $ and significant fluctuations are observed. They do not vanish when averaging over the initial quasi-momentum $\beta$. At long times $ t = 10 000 = 400 T$, the CFS peak at $ x = x_0$ is clearly visible, and duplicates the CBS peak. The fluctuations have also vanished. }
 \label{fig:CBFSvraiQKR}
\end{figure} 

Another aspect that needs to be taken into account for an experimental study of the CBS and CFS peaks with KR systems is the fact that the phases $ e^{-i \alpha_l} $ are not random but given by $ e^{-i \hbar_e (l + \beta)^2/2} $. This means that, in an actual experiment, the spatial distribution $n(x,t)$ is not averaged over the phase disorder but over the initial conditions, {\it i.e.}, over the initial quasi-momenta $ \beta \in [0,1)$. Far from quantum resonances, when $\hbar_e/(2\pi)$ is sufficiently irrational \cite{wimberger2003quantum}, the correlations present in these pseudo-random phases have no dramatic impact on the expansion of a wave packet in momentum space. They essentially renormalize the initial diffusion coefficient \cite{shepelyansky1987localization, tian2010theory}.

However, when considering interference signatures in reciprocal space, such as the CBS and CFS peaks, these correlations have important effects. 
We observed that they affect both the dynamics of the CFS contrast and the spatial distributions of the CBS and CFS peaks. We have not systematically studied these fluctuations and leave this for further studies. We rather provide here a means of mitigating these effects. 

A way to circumvent these fluctuations is to consider the following modulated KR Hamiltonian:
\begin{equation}\label{eq:HKRmod}
 H_\text{mod}=\frac{p^2}{2} - K (1 + \varepsilon \cos\omega_2 t) \cos x \sum_n \delta(t-n) 
\end{equation}
where $\varepsilon$ is the modulation amplitude and $\omega_2$ the angular modulation frequency. The motivation for this choice is twofold. First, a \textit{periodic} modulation of the kick amplitude (with period $T$) allows to simulate a quasi-1D configuration with $T$ channels
\footnote{Note that if $ \omega_2$ is incommensurable with $ 2 \pi $, the KR becomes quasi-periodic in time and its dynamics is effectively 2D \cite{casati1989, chabe2008experimental, lemarie2009}.}. 
This allows to increase significantly the localization length $\xi_o \to T \times \xi_o$ without resorting to prohibitively large $K$ values as done in the previous section. 
Second, the fluctuations observed in position outside the CBS and CFS peaks are also significantly reduced because the different channels induce an additional averaging washing out the effects of the correlations of the pseudo-disorder at long times. 
This is particularly apparent in Figure \ref{fig:CBFSvraiQKR}, which shows the time evolution of a wave packet in the space of positions for a set of parameters which is experimentally accessible \cite{ourPRL2016}, namely $ \hbar_e = 2.89$, $ K = 6$, $ \varepsilon =  0.8$, and $\omega_2=2 \pi \frac{64}{25} $. With these parameters, the effective Heisenberg time is $\tau_o \approx T=25$. Starting from a wave packet initially peaked at $ x_0 = - \pi / 2 $ (but not taking into account the width of a realistic initial state, see below) one observes a CBS peak appearing at $-x_0 = \pi / 2 $ after a few kick periods ($t = 10^2 = 4 T $). At much longer times ($ t = 10^4 = 400 T $), a CFS peak replicating the CBS peak at $ x_0$ is clearly visible. 

\begin{figure}
\includegraphics[width=0.7\linewidth]{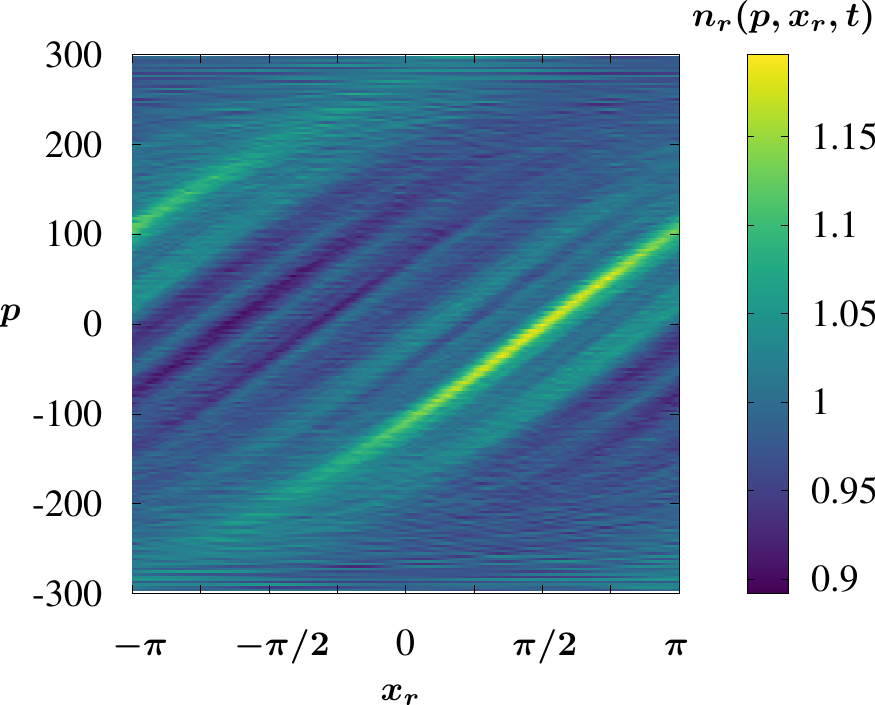}\\
\includegraphics[width=0.7\linewidth]{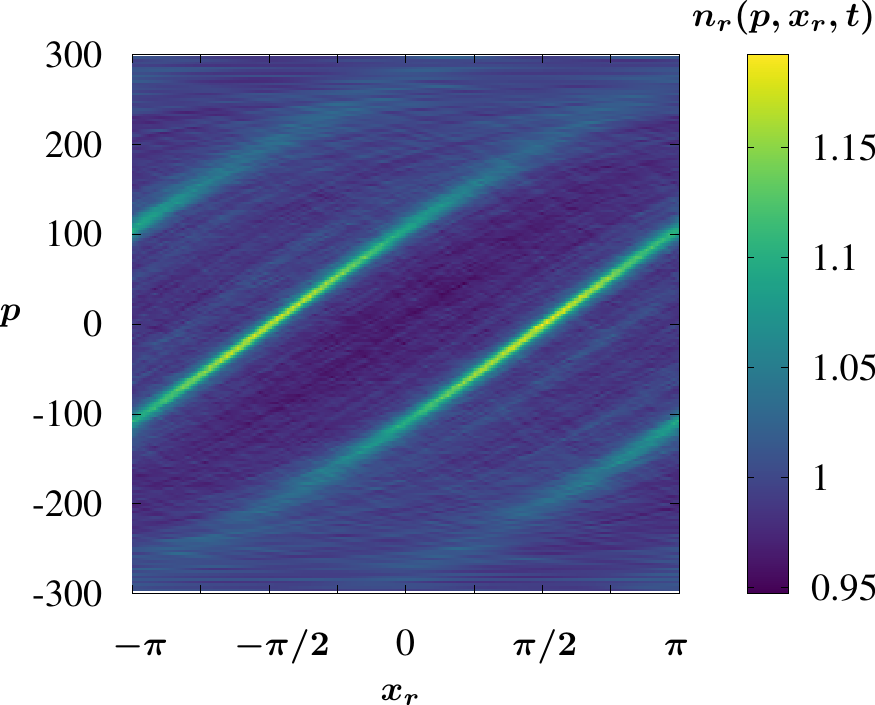}
 \caption{Momentum distribution obtained after phase space rotation as a function of the spatial shift $ x_r$ of the standing wave. The standing wave is applied to the spatial distributions shown in Fig.~\ref{fig:CBFSvraiQKR}. Its amplitude $\gamma=5000$ has been chosen (unrealistically) very large in order to distinguish most clearly the two momentum structures corresponding to the CBS and CFS peaks.
Upper panel: At short times $t=100=4T$, the modulated KR, Eq.~\eqref{eq:HKRmod}, shows a single CBS peak at $x=-x_0=\pi/2$
which manifests itself in the color plot as a single high-amplitude (bright green) line of equation $p\approx\sqrt{\gamma} (x_r+x_0)$. Lower panel: At longer times $t=10^4=400 T$, an additional high-amplitude (bright green) line of equation $p\approx\sqrt{\gamma} (x_r-x_0)$ appears and is a signature of the CFS peak which has emerged at $x= x_0$. 
 }
 \label{fig:rotaCBFSvraiQKR}
\end{figure}

\subsection{Phase space rotation}
\begin{figure}
\includegraphics[width=0.7\linewidth]{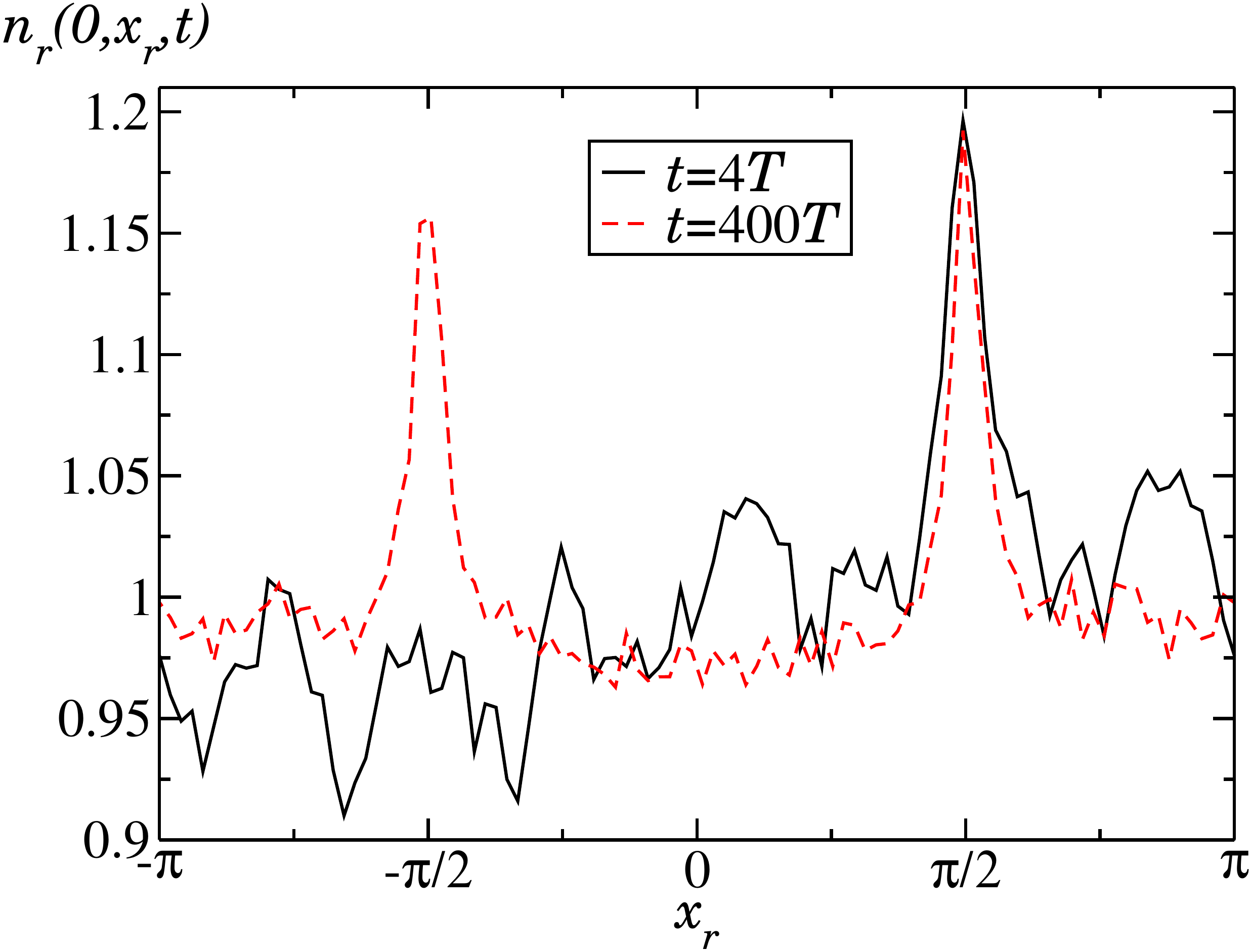}
 \caption{Zero momentum class $n_r(p=0,x_r,t)$ obtained after phase space rotation of the CBS and CFS peaks, as a function of the spatial shift $ x_r$ of the standing wave. Solid black line: momentum class obtained at $t=4T$. Dashed red line: momentum class obtained at $t=400T$. The data correspond to Fig.~\ref{fig:rotaCBFSvraiQKR}. We get a particularly accurate picture of the corresponding spatial distributions shown in Fig.~\ref{fig:CBFSvraiQKR}.
}
 \label{fig:rotaCBFSvraiQKRzero}
\end{figure}

\begin{figure}
\includegraphics[width=0.7\linewidth]{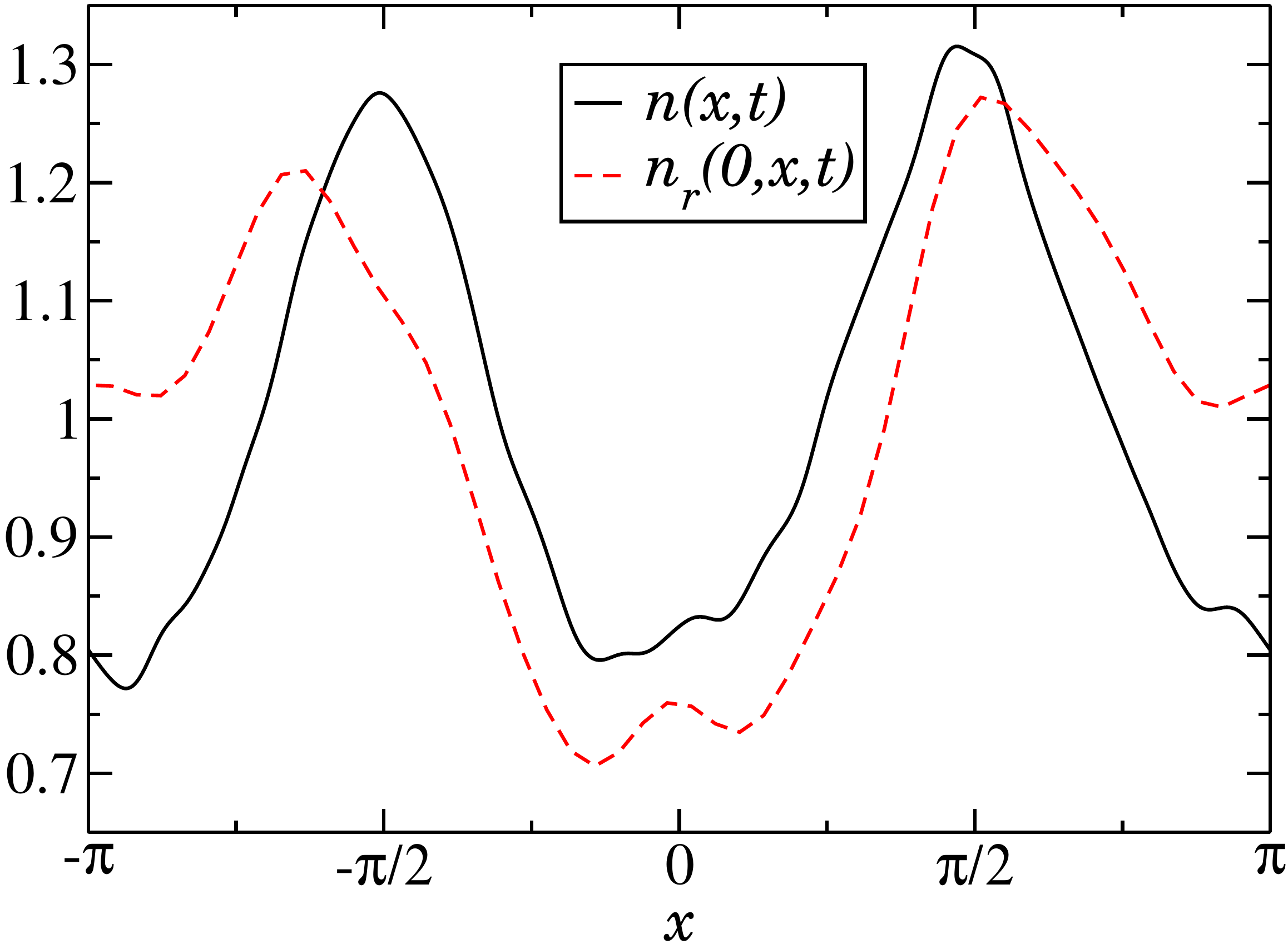}\\
\includegraphics[width=0.7\linewidth]{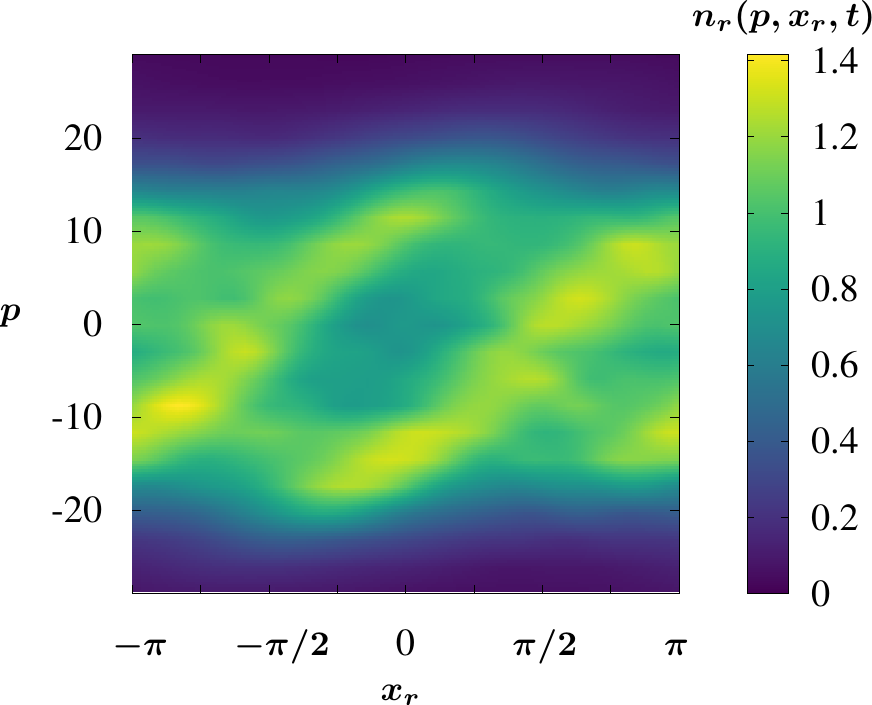}
 \caption{Upper panel: CBS and CFS peaks for the periodic KR, Eq.~\eqref{eq:HKR}, with a realistic initial state. After phase-space rotation, the population of the zero velocity class $n_r(p=0,x_r,t)$ (dashed red line) as a function of the spatial shift $x_r$ shows a good agreement with the averaged spatial distribution $n(x,t)$ (black solid line). Realistic parameters have been chosen to maximise the CBS and CFS contrasts, i.e. $\sigma_x \xi_o <1$ (see Fig.~\ref{fig:inista}), $K=5$, $\hbar_e=2.89$, $t=500$, $\sigma_x\approx0.33$, $\gamma=100$. Lower panel: Momentum distribution after phase space rotation as a function of the phase shift $x_r$. The CBS and CFS peaks at $x=\pm x_0$ in the original spatial distribution are represented by two thick lines of equations $p=\sqrt{\gamma} (x_r\mp x_0)$. 
 }
 \label{fig:trueKR}
\end{figure} 

The spatial distribution of the final state shows a CBS and a CFS peaks of extension $ 1/\xi_o \ll 2 \pi$ along $x$, a pattern which repeats itself along $ x $ with period $ 2 \pi $. While the direct observation of these structures, which have a characteristic size much smaller than the wavelength of the kick potential, is not realistically conceivable, it is possible to rotate these peaks in phase space. If the phase space pattern did not repeat along $ x $, a very natural rotation would be achieved by subjecting the wave packet to a harmonic potential for a quarter of an oscillation period, as in the delta-kick cooling method \cite{ammann1997delta,pmdgo14}. 
Since the phase space distribution for the QKR is periodic, we propose instead to subject the system to a standing wave. Indeed, this standing wave can be approximated by a harmonic potential around each of its minima. The phase space rotation should thus be imprinted identically for each unit cell pattern of the periodic QKR distribution. We therefore consider the evolution of the final wave packet under the influence of a standing wave with the same spatial period as the kick potential. It is described by the Hamiltonian of a pendulum:
\begin{equation}\label{eq:Hrota}
 H_r=  \frac{p^2}{2} - \gamma \cos (x-x_r) \; ,
\end{equation}
where $ \gamma $ is the dimensionless amplitude of the standing wave and $ x_r $ is the spatial offset of the standing wave relative to the stationary kick potential $V_{\text{kick}}(x)$ used in Eq.~\eqref{eq:HKR}. 

Around each of the minima $ x = x_r \; \text {mod} [2 \pi] $, the period of (weak-amplitude) oscillations of the pendulum is $T_r = 2 \pi/\sqrt {\gamma}. $ Let us consider a wave packet localized at $ x = 0 \; \text {mod} [2 \pi] $. Its phase space distribution is a periodic pattern, along the $x$-axis, of lines parallel to the $p$-axis. If we turn on the standing wave during a quarter of the pendulum period $T_r$, the wave packet will start to move and go down the harmonic wells until it reaches its minima at the time when the standing wave is switched off \cite{ourPRL2016}. The wave packet will thus rotate by $\pi/2$.  As a consequence, the resulting phase space distribution is obtained by rotating the initial one by $\pi/2$ and is thus peaked around $ p =0$ and delocalized along the $x$-axis.

We now apply the Hamiltonian Eq.~\eqref{eq:Hrota} onto the final state obtained with the QKR during $T_r/4$ and perform the phase space rotation. We show in Fig.~\ref{fig:rotaCBFSvraiQKR} the result of such a rotation on the CBS and CFS peaks originally obtained in the reciprocal space, see Fig.~\ref{fig:CBFSvraiQKR}. The momentum distribution $n_r(p,x_r,t)$, normalized such that $\int_{-\pi}^{\pi} n_r(0,x_r,t) \, dx_r/(2\pi) = 1$, is represented as a color plot for different values of the spatial offset $x_r$. At small times where only the CBS peak is present, mainly a single line of equation $p\approx \sqrt{\gamma} (x_r+x_0)$ is observed. At large times, a second line is present at $ p\approx \sqrt{\gamma} (x_r-x_0)$ which is a signature of the CFS peak at $x_0$. In particular, in Fig.~\ref{fig:rotaCBFSvraiQKRzero} we plot the momentum density $ n_r(0,x_r,t)$ obtained at zero momentum as a function of the offset distance $x_r$ of the standing wave. We observe two peaks at $x_r = \pm x_0$. Changing the offset $ x_r$ therefore allows to reconstruct in momentum space the two CBS and CFS peaks observed in the spatial distribution. As one can see, the evolution of $ n_r(0,x_r,t) $ as a function of $ x_r $ gives a particularly faithful picture of the spatial distribution. 

\subsection{A realistic set of parameters}

As seen in the previous subsections, the main limitations to an experimental observation of the CBS and CFS peaks are the preparation of the initial state and the amplitude of the stationary wave used to perform the phase-space rotation. In both cases, the important parameter is the depth of the standing wave which is reachable experimentally. 

As discussed previously, well-contrasted peak structures are obtained when $\sigma_x \xi_o < 1$. Since $\sigma_x\approx 0.33$ can be reached experimentally for $s\approx 80$, this limits the possible values of the localization length to $\xi_o < 3$. We therefore chose to work with the periodic KR, Eq.~\eqref{eq:HKR}, at $K=5$ and $\hbar_e = 2.89$ such that the localization length $\xi_o \sim 1$ is small. In Fig.~\ref{fig:trueKR}, we have plotted the CBS and CFS peaks obtained at long times $t=500 \gg  \tau_o$. As one can see they still have a significant contrast. We then rotate these spatial structures with the standing wave Eq.~\eqref{eq:Hrota} with $\gamma=s \hbar_e^2/4 = 100$ (corresponding to $s=50$ for $\hbar_e=2.89$). The zero momentum class after rotation (shown in the upper panel of Fig.~\ref{fig:trueKR}) gives a faithful picture of the original spatial distribution. Note however that the finite width $\sigma_x$ and the limited value of $\gamma$ do not allow to perfectly separate the two structures corresponding to the CBS and CFS peaks, see the lower panels of Fig.~\ref{fig:rotaCBFSvraiQKR} and Fig.~\ref{fig:trueKR} for a comparison.

\section{Conclusion}

With this paper we propose to use the 1D quantum kicked rotor to observe the coherent forward scattering (CFS) peak of strong localization together with its twin brother, the coherent backscattering (CBS) peak of weak localization. 
Whereas CBS has already been observed in numerous systems, a direct experimental observation of CFS has not been reported yet. For systems with spatial disorder, localization occurs in real space and one experimental bottleneck is the dramatic energy spread induced by the disorder on an initial state. On the contrary, localization for the kicked rotor occurs in momentum space and the CBS and CFS structures appear in real space. As the relevant transport observables of the kicked rotor (localization length, Heisenberg time, scattering mean free path, diffusion constant, etc.) are energy-independent, the experimental observation of coherence peaks should be facilitated. Another advantage of the kicked rotor system is that time reversal invariance can be broken by breaking space inversion, which is easily implemented in experiments. This allows to study and compare CFS and CBS in the orthogonal class (time reversal invariant systems) and in the unitary class (systems with broken time reversal invariance). We have analyzed their time dynamics and confronted our numerical results to known theoretical predictions;
in particular, we have conducted the first tests of the CFS peak in the unitary class  and found very good agreement with the predictions of Ref.~\cite{micklitz2014strong}. 
The primary experimental challenge is how to observe narrow peaks in real-space densities at scales significantly smaller than the lattice period, and how to prepare a sufficiently narrow initial state that prevents the coherence peaks to be flattened by convolution. We propose to avoid this bottleneck via a phase-space rotation similar to the delta-kick cooling method, thus transforming peak structures in real space into peak structures in momentum space. We have proposed a realistic experimental protocol to achieve this goal and we conclude that an observation of CBS and CFS with the kicked rotor is within reach of present-day cold-atom experiments. 

As possible future studies to be conducted with the quantum kicked rotor, we foresee CFS and CBS in other universality classes, in particular the symplectic one, CBS and CFS for pseudo-integrable systems and mimicking higher dimensions using frequency modulation. One could then target the 3D Anderson transition in various universality classes and study the impact of multifractal aspects, as observed at the critical point, on CBS and CFS \cite{ghosh2016cfs3d}. Another fascinating perspective would be to probe the effects of interaction on the CBS and CFS peaks \cite{hartung2008coherent, cherroret2014nonlinear, ermann2014destruction, cherroret2015thermalization}.

\ack

GL wishes to thank Nicolas Cherroret, Dominique Delande, Jean-Claude Garreau and Isam Manai for useful discussions. ChM wishes to thank Laboratoire Collisions Agr\' egats R\'eactivit\'e and Laboratoire de Physique Th\'eorique (IRSAMC, Toulouse) for their kind hospitality. CAM acknowledges an invited professorship of Universit\'e Paul Sabatier. We thank CalMiP for access to its supercomputer.
This work was supported by Programme Investissements
d'Avenir under the program ANR-11-IDEX-0002-02, reference  ANR-10-LABX-0037-NEXT and  by  the  ANR  grant
K-BEC  No  ANR-13-BS04-0001-01.

\section*{References}
\bibliography{QKRCBFS_02}

\end{document}